\documentclass[12pt]{article}
\usepackage{amssymb}
\usepackage{amsmath}
\usepackage{booktabs}
\usepackage[hidelinks]{hyperref}
\usepackage{longtable}
\usepackage{graphicx}
\usepackage{subcaption}
\usepackage{ragged2e}
\usepackage{array}
\usepackage[numbers]{natbib}
\usepackage{authblk}
\usepackage{xcolor} % for \textcolor
\newcolumntype{P}[1]{>{\RaggedRight\arraybackslash}p{#1}}
\title{A Multi-Level Data-driven Framework for Understanding Perceptions Towards Cycling Infrastructure Across Regions Leveraging Social Media Discourse}

\author[1]{Shiva Azimi}
\author[2]{Arash Tavakoli}

\affil[1]{Civil and Environmental Engineering, Villanova University, 800 E Lancaster Ave, Villanova, PA 19085, USA \\
Email: \textcolor{blue}{sazimi@villanova.edu}}

\affil[2]{Civil and Environmental Engineering, Villanova University, 800 E Lancaster Ave, Villanova, PA 19085, USA \\
Email: \textcolor{blue}{arash.tavakoli@villanova.edu}}

\date{}

\begin{document}
\maketitle

% \begin{frontmatter}



%% Abstract
\begin{abstract}

Cycling plays an important role in sustainable urban mobility, yet how people perceive cycling infrastructure varies widely and remains challenging to assess at large spatial scales. Existing research has mainly relied on surveys or short-form social media data and has often focused on individual cities, leaving limited insight into how cycling discussions unfold across broader geographic contexts. This study proposes a multi-scale framework that examines how cycling infrastructure is discussed and evaluated in online public discourse and explores whether sentiment patterns differ between the United States (U.S.) and selected European countries included in the dataset. The analysis draws on a large collection of discussions on a social media platform, namely Reddit, including more than 30,000 posts and over 500,000 associated comments gathered from cycling-focused and geographically defined communities across multiple U.S. states and selected European countries. Using a combination of sentiment analysis, topic modeling, aspect-based classification, and hierarchical statistical modeling, the study evaluates the emotional tone and thematic structure of these discussions and how they vary spatially. Overall sentiment toward cycling is positive in both regions, with slightly higher values observed in the European sample, although differences remain modest. Sentiment tends to become more critical in comment discussions compared to original posts. Topic and aspect analyses show that sentiment is primarily associated with experience-based themes, with most variation occurring within cities rather than between regions. Together, these findings illustrate how discussion-based online data can complement traditional approaches to understanding public perceptions of cycling infrastructure in sustainable urban contexts.

\end{abstract}


% \begin{highlights}

% \item Presents a multi-level framework integrating sentiment, topic, and aspect analysis of cycling infrastructure perception in social media discourse

% \item Applies the framework to large-scale Reddit discussions across multiple U.S. states and selected European countries

% \item Finds that variation in cycling discourse is driven more by discussion themes and infrastructure aspects than by geographic scale

% \end{highlights}

%% Keywords
% \begin{keyword}
% %% keywords here, in the form: keyword \sep keyword
% cycling infrastructure \sep Reddit \sep sentiment analysis \sep VADER \sep SBERT \sep HDBSCAN \sep urban mobility
% %% PACS codes here, in the form: \PACS code \sep code
% %% MSC codes here, in the form: \MSC code \sep code
% %% or \MSC[2008] code \sep code (2000 is the default)

% \end{keyword}

% \noindent\textbf{Keywords:} cycling infrastructure; Reddit; sentiment analysis; VADER; SBERT; HDBSCAN; urban mobility
% \end{frontmatter}
\paragraph{Keywords:} cycling infrastructure; Reddit; sentiment analysis; VADER; SBERT; HDBSCAN; urban mobility

%% Main text
%% =========================

\section{Introduction}
Cycling has become an increasingly important element of sustainable urban mobility strategies, offering environmental, health, and social benefits compared to other means of transportation, especially vehicle-dependent transport systems \citep{PucherBuehler2012,Fishman2016}. However, despite its recognized advantages, perceptions of cycling infrastructure and related safety and comfort concerns vary substantially across populations and contexts \citep{Braun2025,FrielWachholz2025,DelclosAlio2024}.

In some metropolitan areas, investment in protected bike lanes and multimodal transport integration has led to growing public enthusiasm for cycling, whereas in others, inadequate infrastructure and safety concerns continue to hinder adoption \citep{PucherBuehler2012,Handy2014}. Prior research suggests that one contributing factor to this gap between infrastructure provision and cycling uptake is \textbf{how cycling environments are perceived, particularly with respect to safety, comfort, and everyday usability} \citep{Braun2025}. Infrastructure characteristics have also been widely examined through crash-based safety analyses that identify roadway features associated with increased risk for vulnerable road users \citep{chen2025influence,azimi2024investigating}. While such studies provide essential insight into safety outcomes, they primarily capture conditions once incidents occur, leaving open the question of how potentially problematic infrastructure is experienced and perceived in everyday cycling contexts. Understanding these differences in perception is therefore crucial, as attitudes toward cycling are shaped not only by the physical presence of infrastructure but also by how citizens experience, discuss, and collectively frame biking in their communities \citep{blitz2021does,biassoni2023choosing}.

Social media platforms have become powerful tools for understanding public sentiment toward transportation and urban infrastructure. Unlike traditional surveys that are often limited by geographic coverage, small sample sizes, and response bias, social media provides a continuous and diverse stream of user-generated content reflecting real-world experiences \citep{Kitchin2014,lock2020social,li2019mining}. Among these platforms, \textit{Reddit} stands out for its distinctive structure of community-based forums known as \textit{subreddits,} where discussions are organized around specific topics, cities, or regions. Users share first-hand experiences and opinions about local transportation policies, safety incidents, and infrastructure developments. Additionally, Reddit's community-based discussions and long-form comments capture detailed perceptions and debates that go beyond the brevity of other social media platforms \citep{Proferes2021}. Such discussions complement traditional survey methods, offering real-time insights into how people emotionally and thematically engage with urban cycling \citep{Gil2012}.

The following study develops a \textbf{scalable cross-region analytical framework that integrates sentiment analysis, semantic topic modeling, aspect-based classification, and hierarchical statistical modeling within a unified pipeline to systematically characterize how cycling infrastructure is perceived and discussed}. The framework converts large volumes of online discourse into interpretable indicators of emotional tone, thematic concerns, and spatial variation in perception. We apply this framework to cycling-related Reddit discussions across cities and regions in the United States and Europe, demonstrating how user-generated content can be translated into quantitative, policy-relevant evidence about infrastructure experiences. Our work paves the way for methodological approaches in transportation analytics that integrate natural language processing (NLP) and data-driven social analysis to provide scalable and timely insights into mobility behavior and infrastructure perception \citep{Kitchin2014,HuLiu2004,Shirgaokar2021,Lock2020,Chowdhury2023}.

\section{Background and Literature Review}
\subsection{Perception-Driven Barriers to Cycling}
Cycling infrastructure plays a central role in shaping urban mobility and sustainability outcomes. Well-designed bicycle networks contribute to lower emissions, improved public health, and increased accessibility, offering a viable alternative to car-oriented systems \citep{PucherBuehler2012,Handy2014}. However, the perceived safety and comfort of cycling routes often determine whether individuals choose to bike, sometimes more than objective infrastructure quality itself \citep{Aldred2018,Heinen2010}. Perception-driven barriers such as fear of traffic, lack of route continuity, or exposure to unsafe intersections can significantly suppress cycling uptake even in cities with developed infrastructure \citep{Garrard2008,Winters2011}. These perceptions also vary widely across geographic and cultural contexts, reflecting differences in planning practices, street design, and social norms surrounding cycling \citep{piatkowski2019carrots}. Understanding how people interpret and emotionally respond to cycling conditions is therefore critical for effective transportation planning, since these perceptions mediate the link between infrastructure investment and behavioral change \citep{Fishman2016}. 

Moreover, recent literature highlights that attitudes and social perceptions toward cycling significantly influence travel choices and interact with physical infrastructure and social environments, shaping experiences of safety, comfort, and accessibility \citep{Willis2015,Banerjee2022}. As cities expand their cycling networks, analyzing both the physical environment and the discursive perception of cycling has become essential for comprehensive infrastructure evaluation. Understanding how these perceptions are measured has therefore become a central focus of transport research, traditionally approached through surveys and stated preference studies that capture how people think and feel about cycling in specific contexts.
% \vspace{0.5em}

\subsection{Traditional Approaches to Cycling Perception Research}
Surveys, interviews, and stated or revealed preference studies have long formed the foundation of cycling perception research, allowing planners to understand how people evaluate safety, comfort, and route choice \citep{StinsonBhat2003}. These methods have been instrumental in connecting subjective attitudes with observed travel behavior, offering interpretable evidence for policy and infrastructure design. Empirical research has shown that perceived safety and convenience strongly influence commuting decisions \citep{StinsonBhat2003}, and that attitudes, habits, and built environment factors jointly shape cycling behavior \citep{Heinen2010}. Other studies emphasize that social context and perceived safety play a key role in determining cycling adoption \citep{Willis2015}.  
% \vspace{0.5em}

Although these traditional approaches remain valuable for their interpretability, they face persistent limitations. Surveys are costly and infrequent, restricting their spatial and temporal coverage \citep{Fishman2016}, and they often suffer from recall and desirability biases \citep{Winters2011}. Attitudinal data also tend to capture static perceptions that fail to reflect the dynamic evolution of cycling environments \citep{handy2014promoting}. Recognizing these constraints, recent work has begun to explore alternative data sources capable of capturing public perception more continuously and across larger geographic and temporal scales.
% \vspace{0.5em}

\subsection{Application of Social Media Analytics in Cycling Perception Research}
In recent years, the increasing availability of digital traces and social media data has transformed how researchers study public perceptions of transportation systems. These data sources complement traditional survey methods by capturing spontaneous, unsolicited, and real-time expressions of opinion \citep{Kitchin2014}. Unlike surveys, which are often limited by small samples and infrequent collection \citep{das2023leveraging,alam2021identifying,yang2024intangible}, social media platforms provide continuous and large-scale observations that reflect lived experiences across diverse urban contexts. Studies examining Twitter discussions about protected bike lane installations have shown that online conversations can reveal public sentiment and recurring themes in response to infrastructure interventions \citep{Ferster2021}. Similarly, research on micromobility systems such as bike-sharing demonstrates how social media content reflects both enthusiasm and criticism toward emerging mobility technologies \citep{DuranRodas2020}. Broader reviews of crowdsourced and digital data in transport research further highlight their potential to complement conventional data sources by offering fine-grained temporal and spatial insights into cycling behavior and perception \citep{Nelson2021}.  
% \vspace{0.5em}

Recent studies have begun applying social media analytics to explore mobility perceptions and public attitudes toward cycling infrastructure. Twitter has been the most frequently used platform in this emerging field, serving as a source of spontaneous and location-specific discussions about safety, policy, and urban design. Analyses of Twitter discourse have captured public reactions to protected bike lane projects and identified shifts in sentiment over time \citep{Ferster2021}. Other work has examined the global discussion surrounding bike-sharing systems, revealing both enthusiasm for sustainable mobility and concerns about maintenance, equity, and urban clutter \citep{DuranRodas2020}. Broader applications of social media data in transportation research have extended to topics such as travel satisfaction, pedestrian safety, and micromobility acceptance, illustrating the versatility of user-generated content for understanding mobility behavior \citep{Nelson2021}.  
% \vspace{0.5em}

While these studies demonstrate the promise of social media for transport perception analysis, most remain limited in scope. They typically focus on single cities or specific modes, employ descriptive sentiment analyses, and lack hierarchical or cross-regional comparisons. Few have examined multi-level interactions between posts, comments, and geographic context, particularly across continents. These gaps motivate the present study, which uses Reddit to provide a broader, comparative view of cycling perception across the United States and Europe.
% \vspace{0.5em}

Most studies employing social media data in transportation research rely on sentiment analysis and topic modeling to extract meaning from large text corpora. Sentiment analysis provides a quantitative measure of emotional tone, typically distinguishing between positive, negative, and neutral expressions. Lexicon-based tools such as the Valence Aware Dictionary for Sentiment Reasoning (VADER) are widely used for social media text because they handle informal language, punctuation, and emphasis effectively \citep{HuttoGilbert2014}. Topic modeling techniques, most commonly Latent Dirichlet Allocation (LDA), have been used to identify dominant discussion themes in transportation discourse, such as safety, infrastructure quality, and mobility policy \citep{Blei2003}. More recent work integrates deep learning and contextual embeddings to improve semantic accuracy in topic discovery. Transformer-based models such as Sentence-BERT and clustering approaches like BERTopic enable the detection of nuanced, context-dependent themes in user-generated content \citep{ReimersGurevych2019,Grootendorst2022}.  
% \vspace{0.5em}

Despite these advances, prior studies often treat sentiment and topics as separate analytical layers, rarely linking emotional tone to specific discussion themes. They also tend to focus on post-level analyses without considering conversational depth or geographic variation. Building on these methods, the present study integrates sentiment and topic modeling within a hierarchical framework that captures post–comment interactions and regional differences in perception.
% \vspace{0.5em}

\subsection{Gaps}
While the use of social media data in transportation research has grown rapidly, several important gaps remain in the literature. First, most studies rely on Twitter, whose short messages limit the depth of discourse and restrict the analysis of complex arguments about infrastructure quality or safety (\textbf{Gap 1}). In contrast, Reddit’s long-form, discussion-based structure allows for multi-level interactions between posts and comments, providing a richer representation of public reasoning and debate. However, Reddit remains largely unexplored as a data source in transport perception studies \citep{Proferes2021}. Second, existing research frequently focuses on single cities or short time frames, which overlooks cross-regional variation in cycling attitudes (\textbf{Gap 2}). Comparative analyses across urban contexts, such as between U.S.\ and European cities, are rare and yet essential for understanding how cultural and infrastructural differences shape perception. Third, while sentiment and topic modeling have become common tools for analyzing mobility-related text, they are typically applied independently (\textbf{Gap 3}). Few studies statistically link emotional tone to specific discussion themes or examine how these relationships vary by geography or conversational depth \citep{Ferster2021,Shirgaokar2021,lock2020social}. Finally, most social media analyses remain descriptive, lacking formal hypothesis testing or hierarchical modeling frameworks that can account for nested data structures (\textbf{Gap 4}).  
% \vspace{0.5em}

To address these gaps, this study introduces a large-scale, cross-continental analysis of Reddit discussions on cycling infrastructure using a multi-level analytical framework. By integrating sentiment analysis with topic modeling, the research captures both the emotional tone and thematic structure of public discourse. Unlike prior studies that focus on single cities or rely primarily on short-form social media data, this study analyzes a substantial corpus of posts and comments spanning multiple cities and regions across the United States and Europe, enabling explicit cross-geographic and cross-continental comparisons of perception patterns.

The framework further incorporates statistical testing and hierarchical modeling to examine how sentiment relates to discussion themes across nested geographic levels. This approach advances existing methodologies by linking emotional tone to thematic content within a structured comparative design. Rather than replacing traditional transport research methods, the study demonstrates how large-scale online discourse can complement them by providing timely, scalable, and geographically diverse insights into how cycling environments are interpreted and discussed. The following section describes the dataset, preprocessing procedures, and analytical methods employed in the study.

\section{Research Questions}

This study investigates how online discussions reflect public perceptions of cycling infrastructure across geographic and thematic contexts. Specifically, we address the following research questions:

\begin{enumerate}
    \item \textbf{RQ1: Regional differences in perception.} 
    How do overall sentiment patterns toward cycling infrastructure differ between the United States and European contexts?

    \item \textbf{RQ2: Thematic drivers of perception.} 
    What dominant themes and infrastructure aspects characterize cycling-related discussions, and how are these themes associated with positive or negative sentiment across regions?

    \item \textbf{RQ3: Spatial scale of variation.} 
    At what geographic scales (country/state versus city) does sentiment variation primarily occur?

    \item \textbf{RQ4: Conversational dynamics.} 
    How does sentiment evolve from original posts to subsequent comment discussions, and does this evolution vary across regions and locations?
\end{enumerate}

% \section{Research Questions}

% This study focuses on the following research questions:

% \begin{enumerate}
%     \item \textbf{RQ1:} How does the sentiment of cycling-related Reddit discussions differ between the United States and the European countries included in the dataset?
    
%     \item \textbf{RQ2:} What dominant themes emerge in cycling-related Reddit discussions, and how do topic prevalence and topic-level sentiment differ across regions?
    
%     \item \textbf{RQ3:} To what extent does city-level variation explain differences in cycling-related sentiment after accounting for higher-level geographic context (state or country)?

%     \item \textbf{RQ4:} How does sentiment vary across cycling infrastructure aspects, and how do these patterns compare between the United States and the European countries included in the dataset?

%     \item \textbf{RQ5:} How does sentiment shift from original posts to the associated comment discussions, and does this post--comment shift differ across regions and geographic units (e.g., U.S. states and European countries)?

% \end{enumerate}

\section{Hypotheses}

We hypothesize the following:

\begin{enumerate}
    \item \textbf{H1:} Post-level sentiment in cycling-related Reddit discussions differs between the United States and the European countries included in the dataset.

    \item \textbf{H2:} Cycling-related sentiment varies across geographic units within each region (U.S. states and European countries).

    \item \textbf{H3:} After accounting for state- or country-level context, cycling-related sentiment exhibits non-zero city-level variation.
\end{enumerate}

\section{Methodology}
This study was reviewed and approved by the Institutional Review Board (IRB) at Villanova University (IRB-FY-2025-76). This project employs a combination of natural language processing (NLP) and aggregation techniques to analyze how cycling infrastructure and safety are discussed across geographic regions in the United States and the European countries included in the dataset for cross-continental comparison. The study focuses on identifying both the emotional tone herein referred to as sentiment, and the main discussion themes expressed in Reddit posts, and on examining how these patterns vary spatially. Below, we describe the dataset (Section~\ref{sec:data}), the sentiment analysis procedure (Section~\ref{sec:sentiment}), the theme extraction approach (Section~\ref{sec:topics}), the aspect-based sentiment analysis procedure (Section~\ref{sec:aspect}), and the spatial comparison framework (Section~\ref{sec:spatial}).

All analyses were implemented in Python~3.10 using open-source libraries including \texttt{VADER}\citep{HuttoGilbert2014}, \texttt{Sentence-BERT}\citep{ReimersGurevych2019}, \texttt{BERTopic}\citep{Grootendorst2022}, and \texttt{HDBSCAN}\citep{McInnes2017}, ensuring reproducibility of the analytical workflow. To support transparency, the scripts implementing the preprocessing and analytical workflow are publicly available in a GitHub repository: \href{https://github.com/shivaazimi7625/Bike_Infrastructure_Perception}{Bike Infrastructure Perception Repository}. The data were collected from publicly accessible Reddit discussions; however, in accordance with Reddit’s platform policies and terms of service, the dataset used in this study is not redistributed. The provided code enables researchers to independently collect comparable data and reproduce the analysis pipeline.

\begin{figure}[t]
    \centering
    \includegraphics[width=0.95\textwidth]{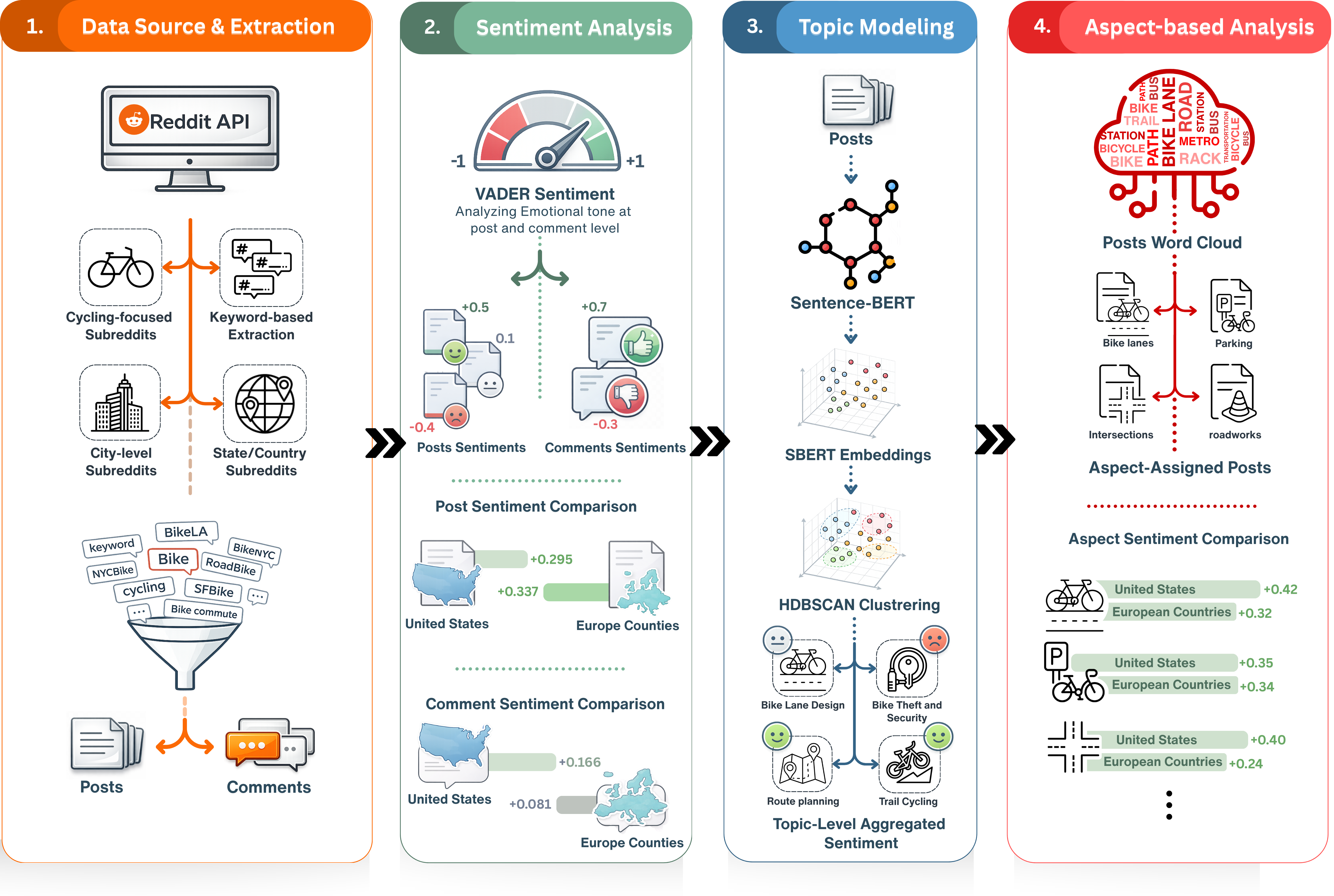}
    \caption{Overview of the data collection, preprocessing, and analysis workflow used in this study.}
    \label{fig:workflow}
\end{figure}

Figure~\ref{fig:workflow} summarizes the overall study workflow and illustrates how the analytical components are connected within a unified pipeline. The process begins with data collection and preprocessing of Reddit discussions, followed by sentiment estimation, topic modeling, and aspect-based classification to capture different dimensions of cycling infrastructure perception. These outputs are then integrated into a multi-level statistical modeling framework to examine variation across geographic contexts. The figure provides an overview of how each methodological stage contributes to the analysis described in the following sections.

\subsection{Dataset Description} \label{sec:data}

Data for this study were collected using three complementary strategies to capture both topic-specific and geographically grounded discussions related to cycling infrastructure, safety, and commuting experiences across the United States and selected European countries. The European subset of the dataset includes posts from the United Kingdom, Germany, France, the Netherlands, Denmark, Spain, Italy, Sweden, Norway, Austria, Belgium, and Switzerland. 

Reddit posts were retrieved using the Reddit API based on predefined subreddit selections and keyword queries. The collected dataset represents posts returned through these queries rather than the complete set of posts within each subreddit. The collected dataset spans discussions between 2010 and 2025 across the United States and selected European countries.

First, posts were gathered from bike-focused subreddits such as \texttt{r/bicycling}, \texttt{r/NYCbike}, and \texttt{r/cycling}, where users frequently share first-hand experiences, opinions, and updates related to biking culture and infrastructure developments. Second, city- and state-level subreddits, including \texttt{r/Seattle}, \texttt{r/Denver}, \texttt{r/LosAngeles}, and \texttt{r/Texas}, were filtered using cycling-related keywords (e.g., \emph{bike}, \emph{lane}, \emph{trail}, \emph{cycling}) to capture locally situated discussions of safety, commuting practices, and road conditions. Finally, additional posts were retrieved through targeted keyword searches that paired cycling-related terms with geographic identifiers, such as ``bike lane in Boston'' or ``cycling in Chicago.'' This final step ensured inclusion of relevant discussions from communities not explicitly focused on cycling but still pertinent to broader mobility and infrastructure topics.

Each record obtained through the Reddit API contains the standard post-level fields provided by the platform, including the post identifier, title, text body, author, creation time, number of comments, score, and upvote ratio. An upvote is a positive signal provided by users to indicate that a post or comment is relevant, useful, or worthy of attention within a community. The score measures community engagement, representing the net number of upvotes minus downvotes that a post receives, while the upvote ratio indicates the proportion of positive votes among total votes (ranging from 0 to 1). Comments associated with each post were also collected and linked using unique identifiers, maintaining the conversational structure of each discussion. Each comment includes a comment ID, textual content, author name, timestamp, and score, allowing the analysis to capture both the primary statements expressed in posts and the subsequent reactions and discussions that follow them.

Table~\ref{tab:posts} presents an excerpt of post-level data showing the original Reddit fields, and Table~\ref{tab:comments} provides a sample of the corresponding comment-level records. The examples are drawn from \texttt{r/NYCbike}, a subreddit focused on cycling-related topics and infrastructure in the New York City metropolitan area.

\begin{table*}[!htbp]
  \centering
  \caption{Excerpt of Post-Level Data (\texttt{r/NYCbike})}
  \label{tab:posts}
  \scriptsize
  \renewcommand{\arraystretch}{1.15}

  \resizebox{\textwidth}{!}{%
  \begin{tabular}{@{}llp{2.8cm}p{5.6cm}lllrl@{}}
    \toprule
    Subreddit & ID & Title & Selftext (excerpt) & Created UTC & Number of Comments & Score & Upvote Ratio & Source Link \\
    \midrule
    r/NYCbike & 1ou8czt &
    3rd Ave Manhattan -- New Bike lane finally underway &
    They just started painting the new lane last night. I don’t know if it goes beyond the newly repaved section from 34th to 42nd Streets yet. &
    2025-10-03 & 8 & 42 & 0.94 &
    \href{https://www.reddit.com/r/NYCbike/comments/1ou8czt/3rd_ave_manhattan_new_bike_lane_finally_underway/}{\texttt{[link]}} \\
    r/NYCbike & 1oc9uh7 &
    Is DOT purposefully building terrible lanes? &
    I have become increasingly convinced that NYDOT is practising malicious compliance with bike lane construction that they are purposefully designing lanes to fail or to frustrate cyclists. &
    2025-09-21 & 24 & 71 & 0.89 &
    \href{https://www.reddit.com/r/NYCbike/comments/1oc9uh7/is_dot purposefully building terrible lanes/}{\texttt{[link]}} \\
    r/NYCbike & 1o5llgx &
    Car driving in protected bike lane crashes into scaffolding in Long Island City &
    Video shows a car entering the protected lane at speed and colliding with scaffolding near Jackson Ave. No injuries reported, but lane was blocked for hours. &
    2025-08-17 & 16 & 58 & 0.91 &
    \href{https://www.reddit.com/r/NYCbike/comments/1o5llgx/car_driving_in_protected_bike_lane_crashes_into/}{\texttt{[link]}} \\
    \bottomrule
  \end{tabular}%
  }
\end{table*}

\begin{table*}[!htbp]
  \centering
  \caption{Excerpt of Comment-Level Data (linked to Table~\ref{tab:posts}).}
  \label{tab:comments}
  \scriptsize
  \renewcommand{\arraystretch}{1.15}

  \resizebox{\textwidth}{!}{%
  \begin{tabular}{@{}llp{10cm}lr@{}}
    \toprule
    Comment\_ID & Parent\_ID & Body (excerpt) & Created\_UTC & Score \\
    \midrule
    c1xv441 & 1ou8czt &
    About time! That section has been terrible for months; finally some smooth asphalt and paint. &
    2025-10-03 & 14 \\
    c1xv556 & 1ou8czt &
    Let’s hope they actually protect it this time. Plastic wands aren’t enough on 3rd Ave. &
    2025-10-03 & 9 \\
    c1xv888 & 1ou8czt &
    It’s crazy how every borough gets a different design. Consistency would make riding safer. &
    2025-10-04 & 11 \\
    \bottomrule
  \end{tabular}%
  }
\end{table*}

In total, the dataset includes approximately 21,000 posts and more than 176,000 comments from cycling-related communities across multiple U.S. states, as well as approximately 10,500 posts and over 330,000 comments from selected European countries included in the dataset. This dataset provides a large and geographically diverse corpus of cycling-related discussions for subsequent sentiment, spatial, and thematic analyses. The complete list of subreddits included in the analysis, along with brief descriptions and subscriber counts at the time of data collection, is provided in~\ref{app:subreddits}. The predefined geographic reference cities used during data collection are provided in ~\ref{app:cities}.

\begin{figure}[t]
    \centering
    \includegraphics[width=0.98\textwidth]{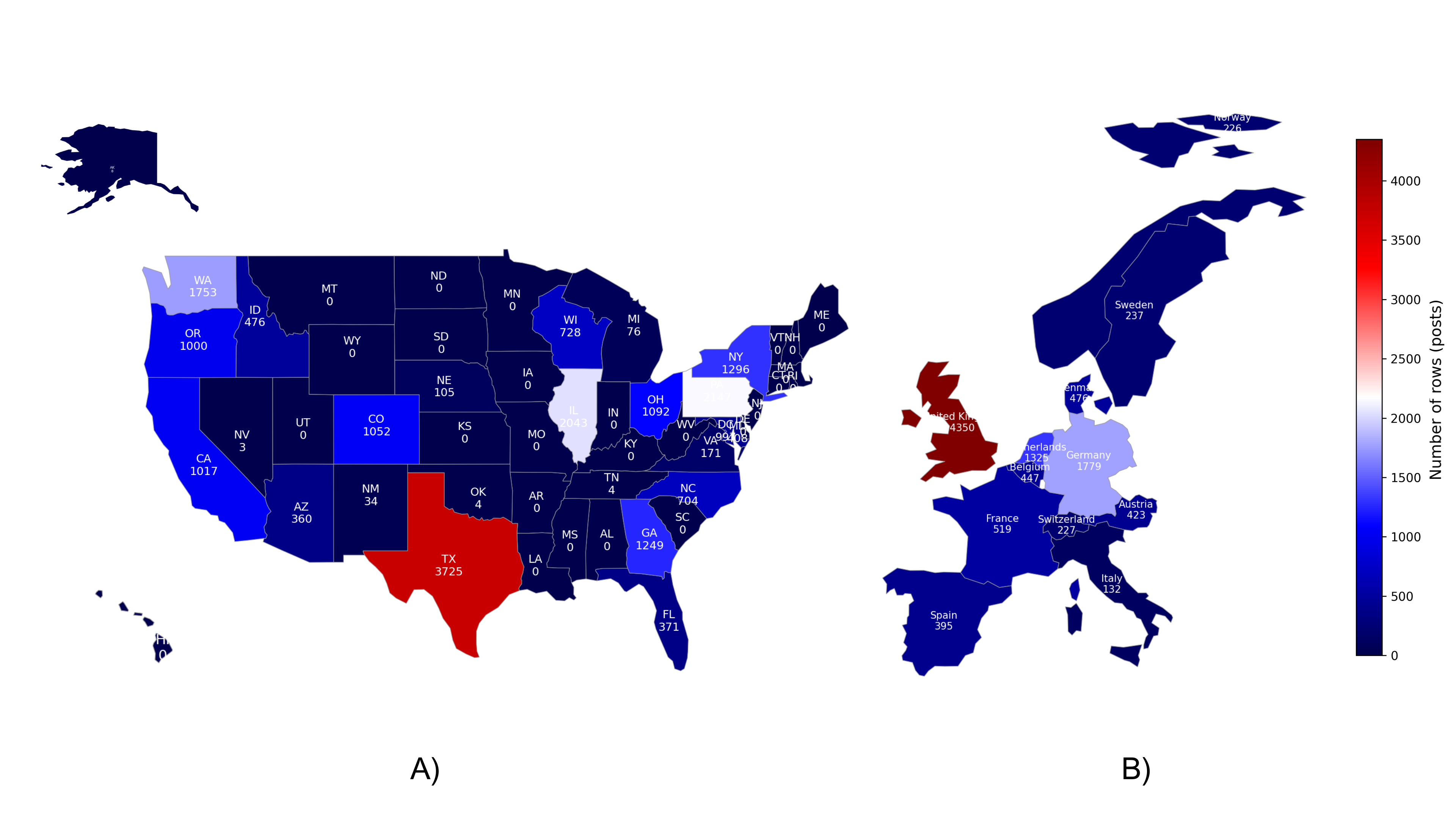}
    \caption{Spatial distribution of Reddit posts in the dataset. 
    (A) Post density across U.S.\ states and (B) post density across European countries. 
    Labels indicate geographic unit names and total post counts. 
    Map geometries are adjusted where necessary to improve visualization clarity and readability.}
    \label{fig:choropleth_spatial_distribution}
\end{figure}

Figure~\ref{fig:choropleth_spatial_distribution} shows the geographic distribution of posts included in the dataset across U.S.\ states and European countries. The figure provides an overview of where cycling-related discussions originate geographically and clarifies the spatial scope of the data used in the analyses that follow.

\begin{table*}[t!]
\centering
\caption{Geographic distribution of Reddit posts across U.S.\ states. The table reports the number of posts, contributing subreddits, and cities represented in each state.}
\label{tab:us_geography}
\setlength{\tabcolsep}{6pt}
\resizebox{\textwidth}{!}{%
\begin{tabular}{lrrp{5.5cm}}
\hline
State & Posts (N) & Subreddits & Cities \\
\hline
Texas & 3725 & 6 & Austin, Dallas, Houston, San Antonio \\
Pennsylvania & 2147 & 4 & Philadelphia, Pittsburgh \\
Illinois & 2043 & 4 & Chicago \\
Washington & 1753 & 3 & Seattle, Tacoma \\
New York & 1296 & 3 & Buffalo, New York, Rochester \\
Georgia & 1249 & 4 & Atlanta, Columbus \\
Ohio & 1092 & 5 & Cincinnati, Cleveland, Columbus \\
Colorado & 1052 & 2 & Colorado Springs, Denver \\
California & 1017 & 2 & Los Angeles, San Diego \\
Oregon & 1000 & 2 & Portland \\
District of Columbia & 997 & 1 & Washington \\
Wisconsin & 728 & 2 & Madison, Milwaukee \\
North Carolina & 704 & 2 & Charlotte \\
Idaho & 476 & 1 & Boise City \\
Maryland & 408 & 1 & Baltimore \\
Nevada & 3 & 1 & Las Vegas \\
Florida & 371 & 4 & Jacksonville, Miami, St.\ Petersburg, Tampa \\
Arizona & 360 & 1 & Phoenix \\
Virginia & 171 & 1 & Richmond \\
Nebraska & 105 & 1 & Omaha \\
Michigan & 76 & 2 & Detroit, Grand Rapids \\
New Mexico & 34 & 1 & Las Cruces \\
Alaska & 8 & 1 & Anchorage Municipality \\
Tennessee & 4 & 2 & Chattanooga, Knoxville \\
Oklahoma & 4 & 2 & Oklahoma City, Tulsa \\
\hline
\end{tabular}%
}
\end{table*}

\begin{table*}[t!]
\centering
\caption{Geographic distribution of Reddit posts across European countries. The table reports the number of posts , contributing subreddits, and cities represented in each country.}
\label{tab:eu_geography}
\setlength{\tabcolsep}{6pt}
\begin{tabular}{lrrp{5.5cm}}
\hline
Country & Posts (N) & Subreddits & Cities \\
\hline
United Kingdom & 4350 & 8 & Birmingham, Edinburgh, Glasgow, London, Manchester \\
Germany & 1779 & 6 & Berlin, Cologne, Frankfurt, Hamburg, Munich \\
Netherlands & 1325  & 6 & Amsterdam, Eindhoven, Rotterdam, The Hague, Utrecht \\
France & 519 & 3 & Paris \\
Denmark & 476 & 3 & Aalborg, Aarhus, Copenhagen \\
Belgium & 447 & 2 & Antwerp, Brussels \\
Austria & 423 & 3 & Graz, Salzburg, Vienna \\
Spain & 395 & 4 & Barcelona, Madrid, Valencia \\
Sweden & 237 & 2 & Malmö, Stockholm \\
Switzerland & 227 & 1 & Zurich \\
Norway & 226 & 3 & Bergen, Oslo, Trondheim \\
Italy & 132 & 1 & Rome \\
\hline
\end{tabular}
\end{table*}

Tables~\ref{tab:us_geography} and~\ref{tab:eu_geography} summarize the geographic composition of the dataset by U.S.\ state and European country, respectively. For each geographic unit, the tables report the number of posts included in the analysis, the number of contributing subreddits, and the cities represented in the data. Together with the spatial maps, these summaries clarify the geographic scope and structure of the dataset used in subsequent analyses.

\subsection{Sentiment Analysis}
\label{sec:sentiment}

The emotional tone of Reddit discussions is quantified using VADER (Valence Aware Dictionary and sEntiment Reasoner), a lexicon- and rule-based sentiment analysis method developed specifically for sentiment detection in social media contexts \citep{HuttoGilbert2014}. In a lexicon-based approach, sentiment is inferred by matching words in the text to a predefined dictionary in which terms are assigned sentiment scores \citep{liu2012sentiment}. VADER estimates sentiment by combining a manually curated lexical dictionary of sentiment-laden terms with a set of empirically validated linguistic rules that adjust sentiment scores based on contextual features. These rules account for phenomena commonly observed in informal online text, including negation (e.g., ``not good''), intensity modifiers (e.g., ``very'', ``extremely''), capitalization, punctuation, and emphasis through repeated characters or symbols.

VADER is well suited for this analysis because it models sentiment intensity rather than relying solely on categorical polarity labels. As described by \citet{HuttoGilbert2014}, each token in a text contributes to an overall sentiment valence, which is then normalized to produce a compound sentiment score ranging from $-1$ (most negative) to $+1$ (most positive). This compound score has been shown to correlate well with human judgments of sentiment in short, informal text and is widely used as an aggregate measure of emotional polarity in social media analysis. In this study, the compound score is used as the primary sentiment indicator for both posts and comments.

VADER is particularly well suited for Reddit data, where discussions are often conversational, concise, and rich in emphasis, slang, and nonstandard grammar. Unlike supervised machine-learning approaches, VADER does not require domain-specific training data, which supports consistent sentiment estimation across geographically and culturally diverse regions. 

Sentiment scores are computed at the individual post and comment levels and subsequently aggregated across geographic units, including cities, states, and countries, as well as across subreddit communities. This aggregation enables systematic examination of spatial patterns in sentiment and comparison between post-level sentiment and the sentiment expressed within associated comment threads, capturing how collective discussion may reinforce, moderate, or shift the emotional tone of the original content.

\subsection{Theme Extraction}
\label{sec:topics}

To understand \emph{what} issues and concerns people discuss in relation to cycling infrastructure, we extracted discussion themes using BERTopic \citep{Grootendorst2022}. BERTopic is an unsupervised topic modeling framework that groups semantically similar documents into coherent clusters and summarizes each cluster using representative keywords, producing interpretable topics from large-scale text data.

The BERTopic pipeline begins by converting each Reddit post into a numerical representation using Sentence-BERT (SBERT) \citep{ReimersGurevych2019}. SBERT maps each post to a fixed-length vector, commonly referred to as an \emph{embedding}, where texts with similar meaning are positioned closer together in the embedding space. This representation captures semantic similarity at the sentence level, enabling posts that describe the same idea using different wording (e.g., ``protected bike lane'' versus ``separated cycling path'') to be grouped together. This is particularly useful for Reddit data, where users frequently employ informal language, abbreviations, and diverse phrasing.

Posts are then grouped into topics using Hierarchical Density-Based Spatial Clustering of Applications with Noise (HDBSCAN) \citep{McInnes2017}. HDBSCAN identifies clusters by locating dense regions of the embedding space, meaning that posts discussing similar cycling-related issues tend to form concentrated groups. Unlike clustering approaches that require the number of clusters to be specified in advance, HDBSCAN determines an appropriate number of clusters directly from the data. It also labels posts that do not clearly belong to any cluster as outliers, which helps reduce noise and improves the interpretability of the extracted themes.

After clustering, BERTopic generates topic representations by identifying the most representative terms within each cluster, which are used to summarize and interpret the extracted themes. The resulting topics capture recurring themes in cycling-related discussions, including safety concerns, infrastructure quality and design, maintenance conditions, theft, and policy or planning debates. Topic prevalence was computed as the frequency of posts assigned to each topic, and topic-level sentiment was computed by aggregating sentiment scores of posts within each topic. These outputs enable comparison of dominant themes and their associated sentiment across the U.S. and the European countries included in the dataset.

\subsection{Aspect-Based Sentiment Analysis}
\label{sec:aspect}

To examine sentiment toward specific components of cycling infrastructure, we applied a rule-based, keyword-driven aspect-based sentiment analysis, a commonly used approach in aspect-based opinion mining when labeled training data are unavailable \citep{HuLiu2004,liu2012sentiment}. Infrastructure aspects were defined inductively through exploratory analysis of the text corpus. Word frequency tables were generated from cycling-related Reddit posts, and frequently occurring infrastructure-related terms were identified. Related terms were then grouped into a set of infrastructure aspects through iterative inspection to capture common infrastructure-related themes in the corpus.

The final set of aspects included \textit{protected lanes, painted lanes, general bike lanes, paths and trails, parking and storage, intersections and signals, transit integration, construction and roadworks, and general infrastructure.} For each aspect, representative keywords were identified from the corpus and used to assign posts to infrastructure components.

Aspect assignment was conducted at the post level using keyword-based pattern matching. Prior to matching, text was converted to lowercase and punctuation was removed. A post was assigned to an aspect if it contained at least one keyword associated with that aspect. Because posts may reference multiple infrastructure elements, a single post could be assigned to more than one aspect. The same keyword sets and matching procedure were applied consistently across the U.S.\ and European datasets.

Post-level sentiment scores were computed separately using the VADER sentiment analyzer. For each region–aspect combination, sentiment scores were aggregated to produce descriptive summary statistics, including the number of associated posts, mean sentiment, and median sentiment. These summaries are reported in the Results section.

\subsection{Spatial Aggregation and Comparison}
\label{sec:spatial}
Sentiment scores and extracted themes are aggregated geographically to explore spatial variation in online perceptions of cycling. Summary metrics (e.g., median or mean sentiment, topic prevalence) are computed for each city, state, and subreddit to capture both the general tone of discussions and the prominence of specific issues. This comparative analysis highlights where discussions are more positive versus dominated by concerns such as safety hazards or poor design.

\section{Results}
\subsection{Exploratory Word Cloud Analysis}

\begin{figure}[t]
    \centering
    \begin{subfigure}[t]{0.48\textwidth}
        \centering
        \includegraphics[width=\linewidth]{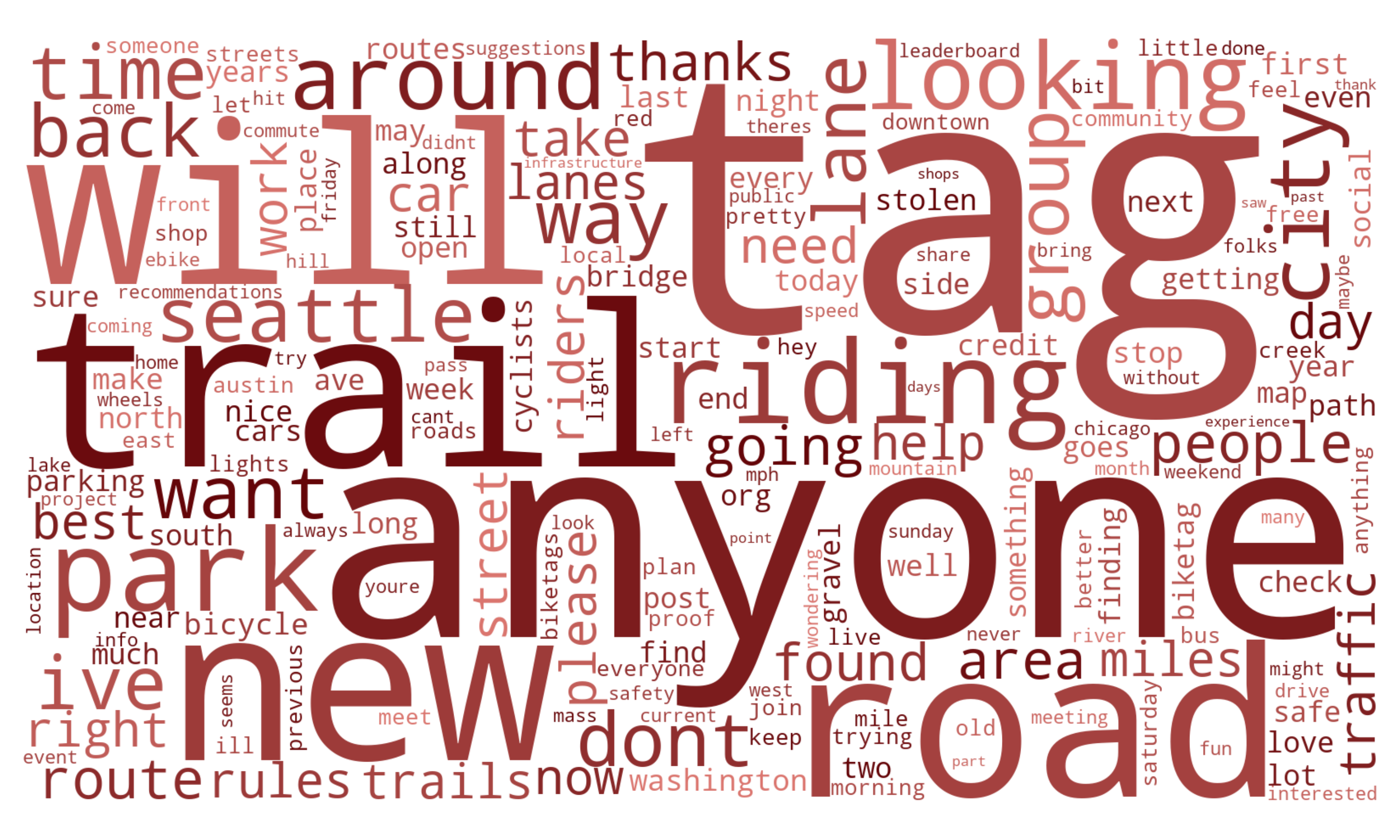}
        \caption{United States}
        \label{fig:wordcloud_us}
    \end{subfigure}
    \hfill
    \begin{subfigure}[t]{0.48\textwidth}
        \centering
        \includegraphics[width=\linewidth]{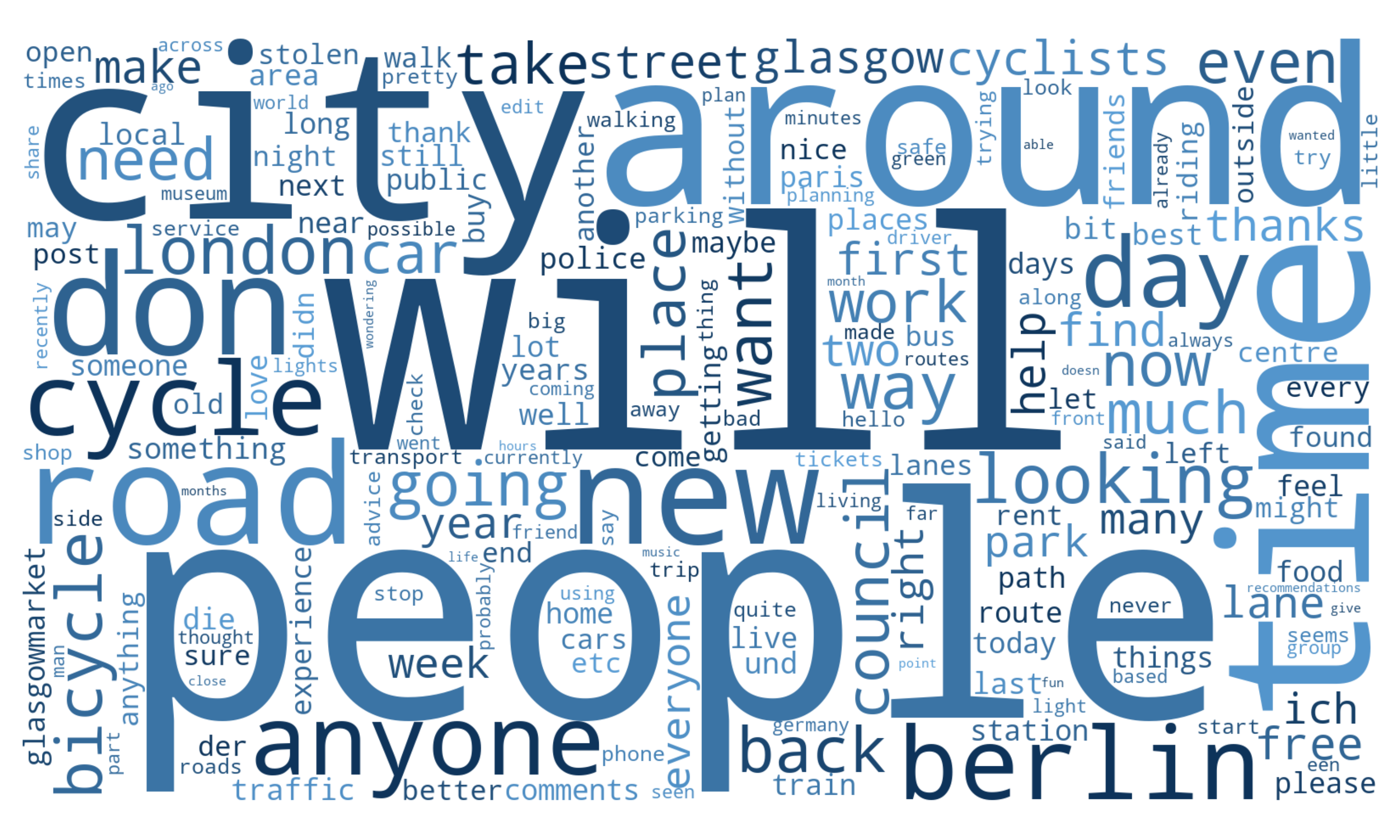}
        \caption{Europe}
        \label{fig:wordcloud_europe}
    \end{subfigure}
    \caption{Word clouds of cycling-related Reddit posts in the United States and Europe. Word size reflects relative term frequency after text preprocessing and stopword removal.}
    \label{fig:wordclouds_us_eu}
\end{figure}

Figure~\ref{fig:wordclouds_us_eu} provides an exploratory overview of the most frequently occurring terms in cycling-related Reddit posts from the United States and Europe. The word clouds highlight commonly used vocabulary in cycling-related discussions and serve as a descriptive complement to the spatial distribution shown in Figure~\ref{fig:choropleth_spatial_distribution}. Differences in term prominence across regions reflect variation in discussion emphasis.

\subsection{Topic Modeling of Cycling Discussions}

\subsubsection{Dominant Topics in Cycling Discussions}

Topic modeling reveals that cycling-related discussions in both the United States and Europe are organized around a small set of recurring themes that reflect how cycling is experienced and discussed in everyday contexts. Separate BERTopic models were estimated for each region, yielding ten dominant topics per region after excluding outlier posts. Table~\ref{tab:topics} summarizes the most frequent topic clusters, their representative keywords, and the number of posts associated with each topic.

Across both regions, discussions related to bike theft and security constitute a prominent component of cycling discourse, indicating that concerns about safety and loss are central to how cycling is discussed online. Other frequently occurring topics are associated with recreational cycling, social or group-based activities, route planning, and travel-related experiences. While several topics appear in both regions, the relative emphasis placed on specific themes differs. In the United States, discussions more frequently center on group rides, commuting, and route planning, whereas European discussions place greater emphasis on place-based cycling cultures, travel experiences, and maintenance or repair activities.

\begin{table}[t]
\centering
\caption{Most frequent topic clusters and associated sentiment in the United States and Europe.}
\label{tab:topics}
\scriptsize
\setlength{\tabcolsep}{4pt}
\renewcommand{\arraystretch}{1.15}

\begin{tabular}{@{}l p{3.4cm} r r p{4.2cm}@{}}
\toprule
Region & Topic description & $N$ & Mean sentiment & Example keywords \\
\midrule
US & Bike theft and security & 711 & -0.17 & stolen, lock, theft, police \\
US & Group rides and social cycling & 445 & 0.52 & group, rides, cycling, join \\
US & Recreational and trail cycling & 296 & 0.39 & gravel, trails, mtb, road \\
US & Bike lane design & 252 & 0.21 & lane, protected, barriers \\
US & Route planning and commuting & 165 & 0.51 & route, commute, safest \\
\midrule
Europe & Bike theft and security & 303 & -0.09 & stolen, lock, theft \\
Europe & Cycling in the Netherlands & 262 & 0.67 & amsterdam, dutch, rent \\
Europe & Bike repair and maintenance & 173 & 0.56 & repair, tools, brakes \\
Europe & Parking and storage & 166 & 0.46 & parking, secure, station \\
Europe & Cycling-related travel & 118 & 0.92 & paris, trip, metro \\
\bottomrule
\end{tabular}
\end{table}

\subsubsection{Sentiment Associated with Cycling Topics}

Sentiment differs across the topic categories identified through topic modeling. In both the United States and Europe, topics related to bike theft and security show the lowest sentiment values, which is consistent with the negative tone of discussions centered on crime, loss, and safety concerns. Topics related to recreational cycling, social activities, and travel experiences tend to show higher sentiment values and are generally discussed in a more positive context.

Differences also appear when comparing how sentiment aligns with topics across regions. In the United States, discussions of group rides, route planning, and recreational or trail-based cycling are associated with higher sentiment values, while topics related to infrastructure design and planning are closer to neutral. In Europe, higher sentiment values are more often observed in discussions tied to place-based cycling cultures and travel experiences in established cycling cities. In both regions, theft-related topics remain the most consistently negative, including discussion clusters that involve multiple languages.

These patterns indicate that differences in aggregate sentiment between regions reflect the types of topics being discussed rather than a uniform change in sentiment across all topics. This observation helps contextualize the regional and geographic sentiment analyses presented in the following sections.

\subsection{Aspect-Based Sentiment Differences}

To investigate how sentiment differs across specific components of cycling infrastructure, an aspect-based sentiment analysis was conducted for the United States and Europe. Reddit posts were categorized into predefined infrastructure-related aspects using keyword-based matching, a commonly used approach for extracting aspect-level information from unstructured text \citep{liu2012sentiment}. Sentiment scores were then aggregated within each region–aspect combination. For each aspect, the total number of posts, mean sentiment, and median sentiment are reported, enabling direct comparison of how different infrastructure elements are perceived across regions.

Table~\ref{tab:aspect_sentiment} summarizes aspect-level sentiment statistics for both regions. In Europe, the highest mean sentiment values are observed for \textit{transit integration} (0.443), \textit{construction and roadworks} (0.399), and \textit{paths and trails} (0.372). In contrast, \textit{bike lanes (general)} exhibit the lowest mean sentiment (0.178). In the United States, the most positive aspects include \textit{general infrastructure} (0.431), \textit{protected lanes} (0.418), and \textit{paths and trails} (0.378), while \textit{painted lanes} and \textit{construction and roadworks} show comparatively lower mean sentiment values.

\begin{figure}[!htbp]
  \centering
  \includegraphics[
    width=0.80\columnwidth,
    height=0.35\textheight,
    keepaspectratio
  ]{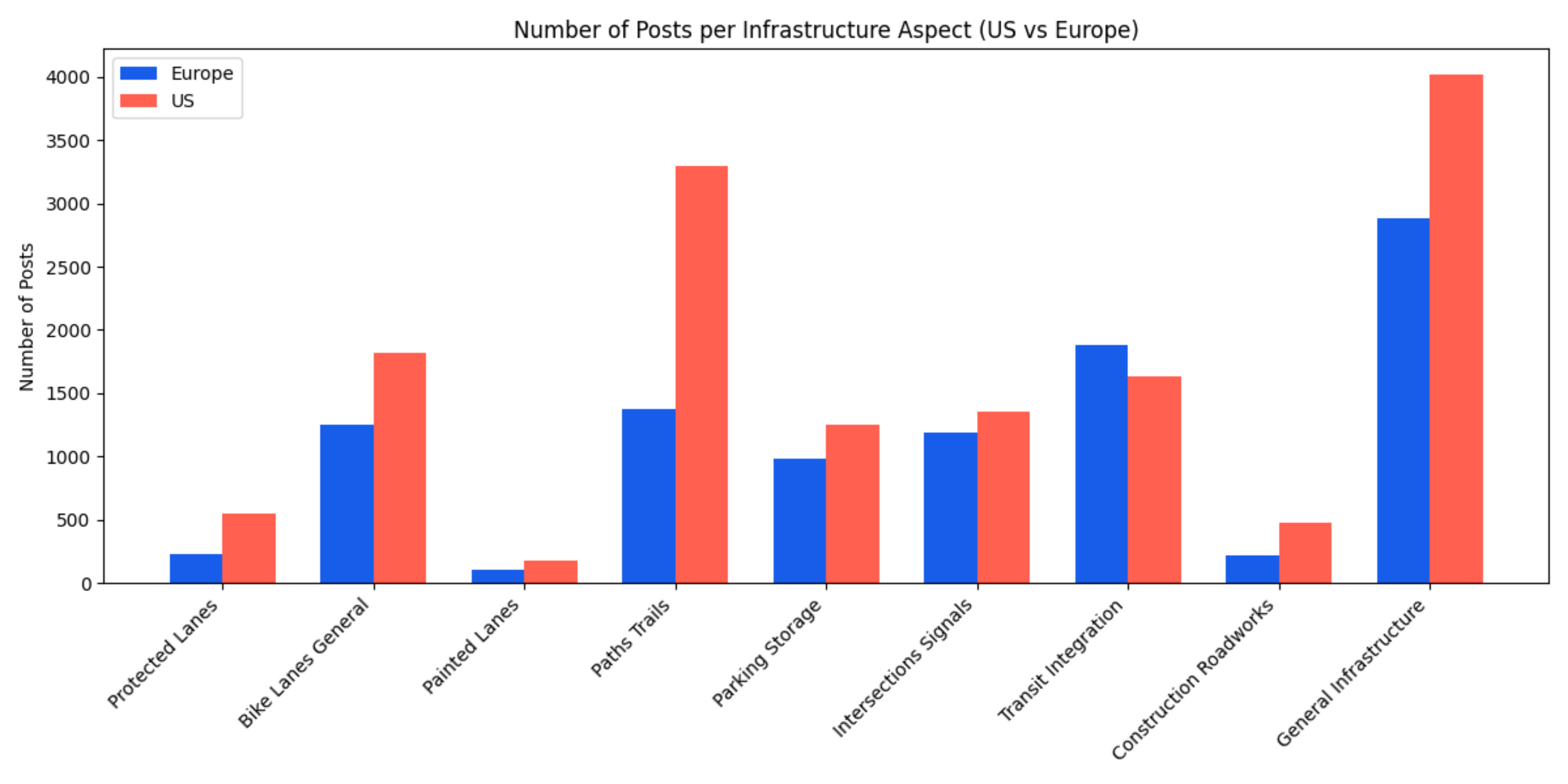}
  \caption{Number of cycling-related posts by infrastructure aspect in Europe and the United States.}
  \label{fig:aspect_counts}
\end{figure}

\begin{figure}[!htbp]
  \centering
  \includegraphics[
    width=0.80\columnwidth,
    height=0.35\textheight,
    keepaspectratio
  ]{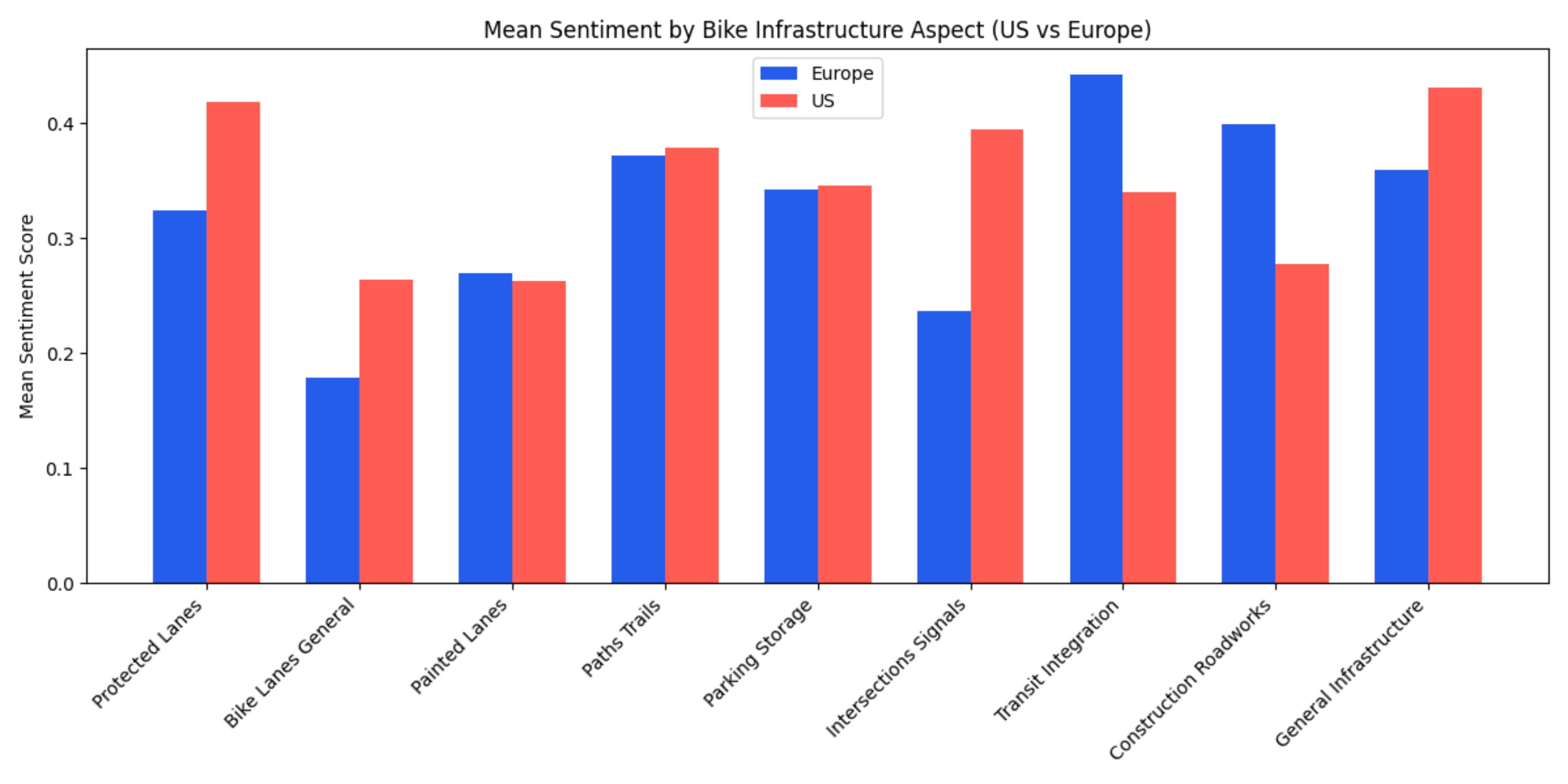}
  \caption{Mean sentiment scores by cycling infrastructure aspect in Europe and the United States.}
  \label{fig:aspect_sentiment}
\end{figure}

Figure~\ref{fig:aspect_counts} shows the number of posts associated with each infrastructure aspect in the United States and Europe. Across both regions, \textit{general infrastructure} and \textit{paths and trails} account for the largest share of discussions, while \textit{painted lanes} and \textit{protected lanes} appear less frequently. Figure~\ref{fig:aspect_sentiment} illustrates mean sentiment scores by aspect and region, highlighting systematic differences in how specific infrastructure components are perceived across the two contexts.

\begin{table}[!htbp]
\centering
\small
\caption{Aspect-based sentiment statistics for cycling infrastructure discussions in Europe and the United States.}
\label{tab:aspect_sentiment}
\setlength{\tabcolsep}{6pt}
\renewcommand{\arraystretch}{1.15}

\begin{tabular}{llcccc}
\toprule
Region & Aspect & $N$ & Mean & Median \\
\midrule
Europe & Protected lanes & 224 & 0.324 & 0.742 \\
Europe & Bike lanes (general) & 1{,}256 & 0.178 & 0.380 \\
Europe & Painted lanes & 107 & 0.269 & 0.462 \\
Europe & Paths and trails & 1{,}371 & 0.372 & 0.697 \\
Europe & Parking and storage & 986 & 0.343 & 0.743 \\
Europe & Intersections and signals & 1{,}185 & 0.236 & 0.579 \\
Europe & Transit integration & 1{,}878 & 0.443 & 0.826 \\
Europe & Construction and roadworks & 216 & 0.399 & 0.880 \\
Europe & General infrastructure & 2{,}883 & 0.359 & 0.745 \\
\midrule
United States & Protected lanes & 544 & 0.418 & 0.599 \\
United States & Bike lanes (general) & 1{,}814 & 0.263 & 0.407 \\
United States & Painted lanes & 172 & 0.263 & 0.502 \\
United States & Paths and trails & 3{,}295 & 0.378 & 0.484 \\
United States & Parking and storage & 1{,}253 & 0.346 & 0.527 \\
United States & Intersections and signals & 1{,}352 & 0.394 & 0.652 \\
United States & Transit integration & 1{,}635 & 0.339 & 0.424 \\
United States & Construction and roadworks & 476 & 0.278 & 0.352 \\
United States & General infrastructure & 4{,}021 & 0.431 & 0.649 \\
\bottomrule
\end{tabular}
\end{table}

\subsection{Sentiment Differences Between Europe and the United States}

We first examine overall sentiment patterns in cycling-related Reddit discussions across the United States and Europe, distinguishing between original posts and subsequent comment discussions.

Table~\ref{tab:sentiment_summary} summarizes the aggregate sentiment statistics for posts and comments in both regions. European posts exhibit a higher mean sentiment score (0.337) than U.S.\ posts (0.295). This difference in post-level sentiment is statistically significant according to the Mann--Whitney U test ($U = 98{,}173{,}614$, $p < 10^{-73}$). The Kolmogorov--Smirnov test further confirms that the overall sentiment distributions differ across regions ($D = 0.186$, $p < 10^{-212}$), indicating systematic differences beyond shifts in the mean alone. 

Despite these statistically significant results, the associated effect size, measured using Cliff’s Delta \citep{Cliff1993}, is close to zero, indicating that the magnitude of the difference between European and U.S.\ post sentiment distributions is small and that the distributions overlap substantially. Given the large sample size, even small distributional differences can reach statistical significance; however, effect size estimates indicate that the magnitude of regional differences remains modest.

In contrast, comment-level sentiment is lower in both regions, with European comments showing a substantially lower mean sentiment (0.081) compared to U.S.\ comments (0.166). Across regions, comments are therefore more negative on average than posts.

In terms of sentiment categories, European posts contain the highest proportion of positive sentiment (65.3\%, $n \approx 6{,}960$), followed by U.S.\ posts (59.7\%, $n \approx 12{,}610$). European comments also exhibit a relatively high proportion of negative sentiment (31.3\%, $n \approx 103{,}600$), compared to 25.5\% ($n \approx 44{,}900$) for U.S.\ comments. The proportion of positive comments is lower in Europe (44.3\%, $n \approx 146{,}600$) than in the United States (51.1\%, $n \approx 89{,}900$), with sentiment decreasing from posts to comments in European discussions.

\begin{figure}[!htbp]
  \centering
  \includegraphics[width=0.75\textwidth]{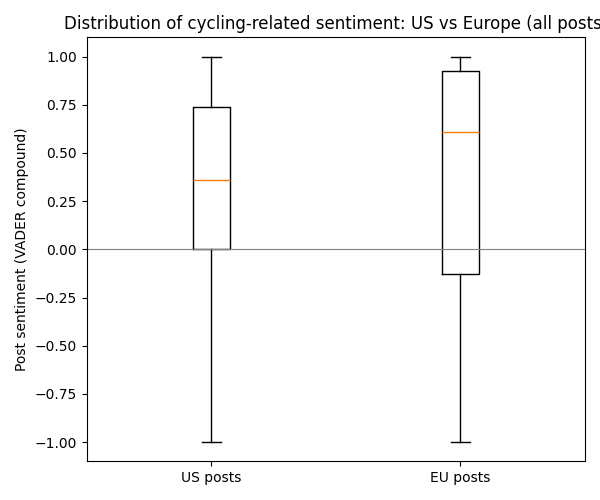}
  \caption{Distribution of post-level sentiment scores for cycling-related discussions in the United States and Europe.}
  \label{fig:sentiment_boxplot}
\end{figure}

\begin{figure}[!htbp]
  \centering
  \includegraphics[width=0.75\textwidth]{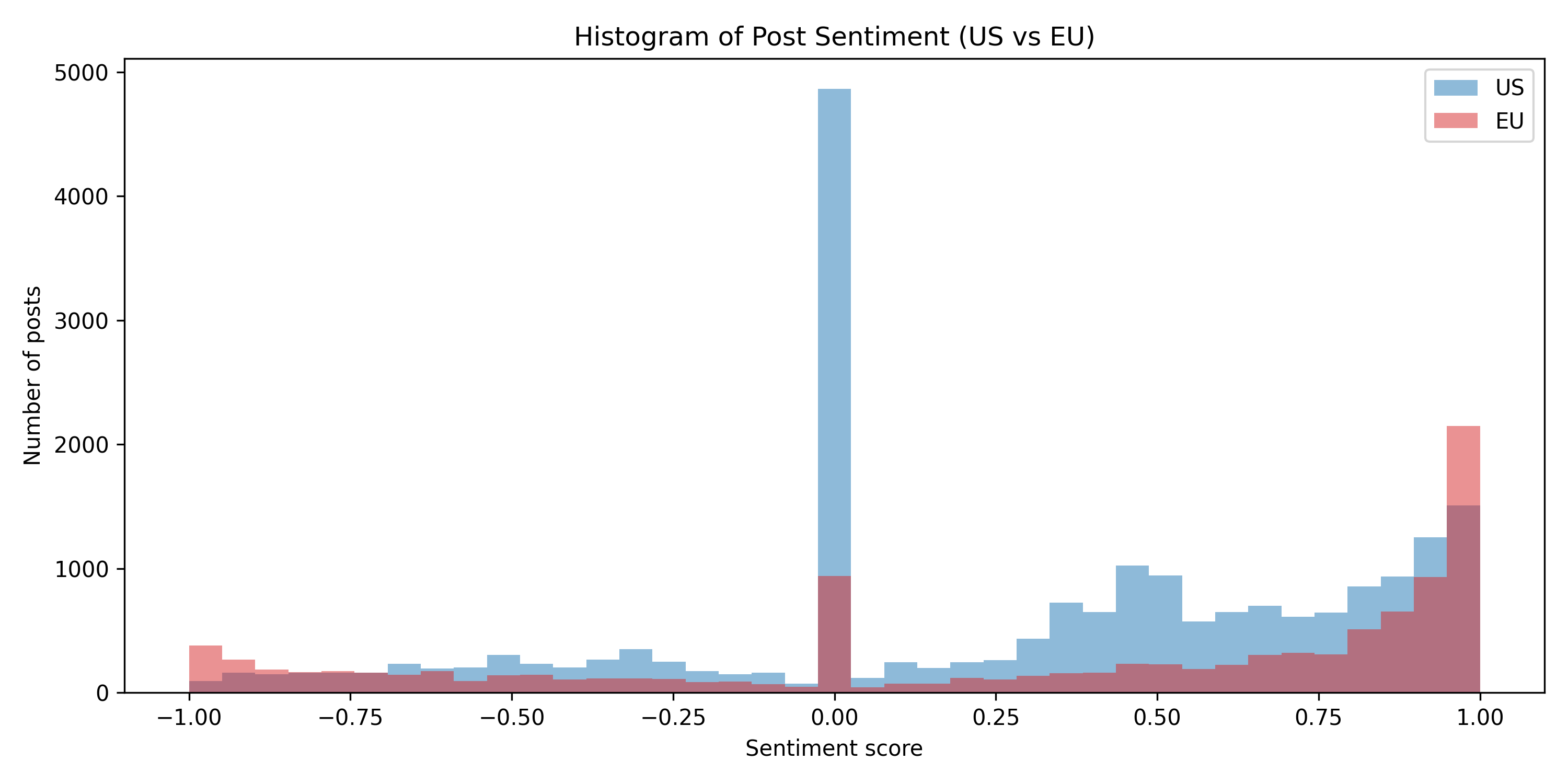}
  \caption{Histogram of post-level sentiment scores for cycling-related discussions in the United States and Europe.}
  \label{fig:sentiment_histogram}
\end{figure}

Figure~\ref{fig:sentiment_boxplot} shows boxplots of post-level sentiment distributions for the two regions. European posts display a higher median sentiment and a wider interquartile range than U.S.\ posts, consistent with the differences identified by the statistical tests. Figure~\ref{fig:sentiment_histogram} further illustrates the distributional differences, with European posts showing a higher concentration of strongly positive sentiment values, while both regions exhibit substantial mass near neutral sentiment.

\begin{figure}[!htbp]
  \centering
  \includegraphics[width=0.75\textwidth]{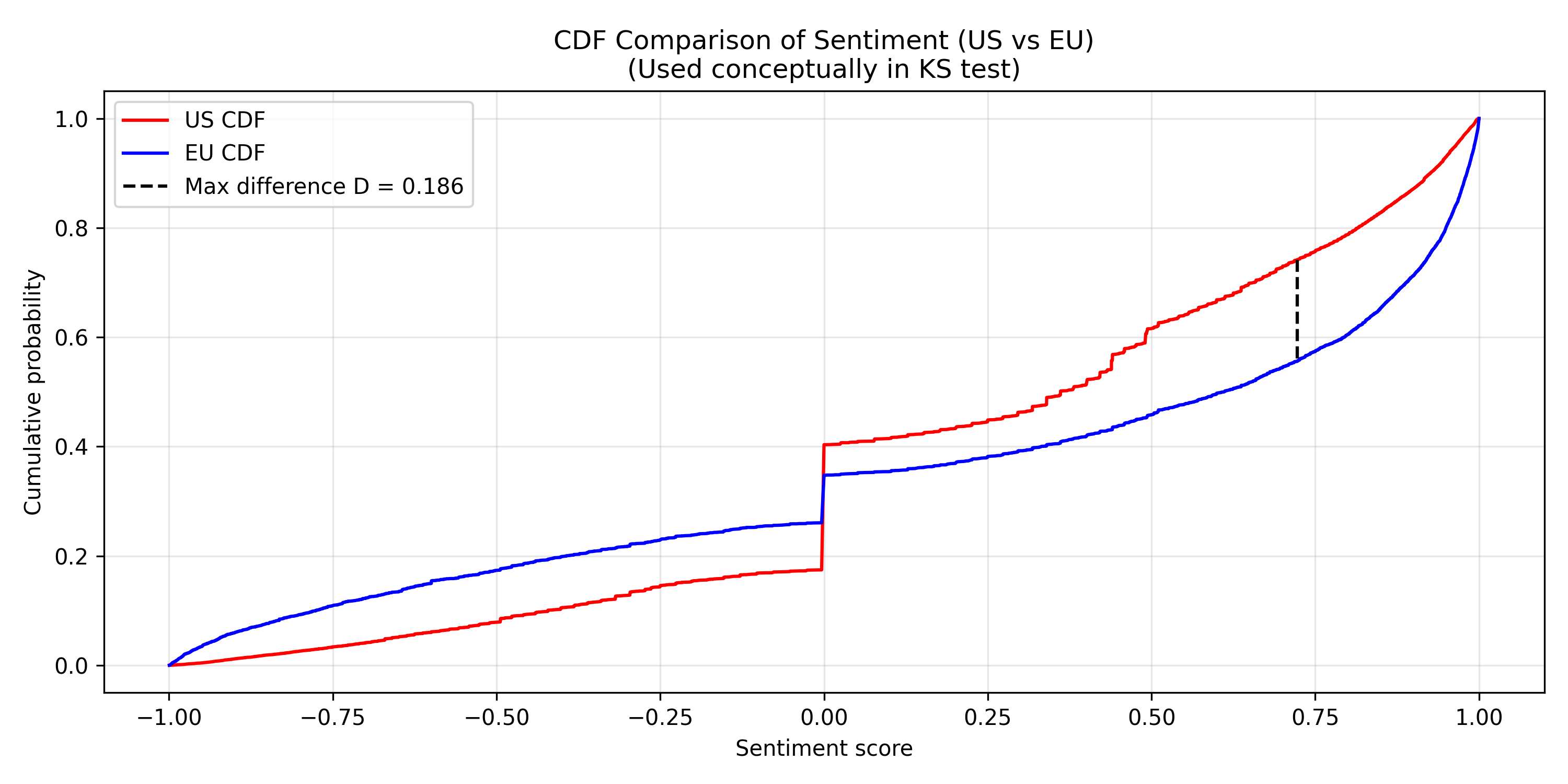}
  \caption{Empirical cumulative distribution functions (CDFs) of post-level sentiment scores for the United States and Europe.}
  \label{fig:sentiment_cdf}
\end{figure}

To compare the full sentiment distributions, Figure~\ref{fig:sentiment_cdf} presents the empirical cumulative distribution functions (CDFs) for U.S.\ and European post sentiment. The maximum vertical separation between the two curves highlights systematic differences across the full range of sentiment values, rather than differences confined to the mean alone.

\begin{table}[h]
\centering
\caption{Aggregate sentiment statistics for cycling-related Reddit posts and comments.}
\label{tab:sentiment_summary}
\begin{tabular}{lcccc}
\hline
Region & Mean Sentiment & Std. Dev. & \% Positive & \% Negative \\
\hline
U.S. Posts     & 0.295 & 0.499 & 59.7 & 17.5 \\
U.S. Comments  & 0.166 & 0.483 & 51.1 & 25.5 \\
EU Posts       & 0.337 & 0.659 & 65.3 & 26.1 \\
EU Comments    & 0.081 & 0.510 & 44.3 & 31.3 \\
\hline
\end{tabular}
\end{table}

% \subsection{Statistical Comparison of Sentiment Distributions Between Europe and the United States}

% To assess whether cycling-related sentiment differs between Europe and the United States, we conducted non-parametric statistical comparisons of post-level sentiment distributions across regions. Differences in central tendency were evaluated using the Mann–Whitney U test \citep{Mann1947}. Distributional differences between regions were examined using the two-sample Kolmogorov–Smirnov test \citep{Smirnov1939}. In addition to hypothesis testing, the magnitude and direction of regional differences were quantified using Cliff’s Delta \citep{Cliff1993}, providing an interpretable measure of effect size for large, non-normally distributed samples.

% The Mann--Whitney U test indicates a statistically significant difference in average post sentiment between the two regions ($U = 98{,}173{,}614$, $p < 10^{-73}$). The Kolmogorov--Smirnov test further confirms that the overall sentiment distributions differ across regions ($D = 0.186$, $p < 10^{-212}$), indicating systematic differences beyond shifts in the mean alone. Levene’s test reveals a statistically significant difference in variance between the two distributions ($F = 974.85$, $p < 10^{-210}$), suggesting differences in sentiment dispersion or polarization.

% Despite these statistically significant results, the estimated Cliff’s Delta values are close to zero, indicating that the magnitude of the difference between European and U.S.\ sentiment distributions is small, with substantial overlap between the two distributions.

\subsection{Post--Comment Sentiment Shifts}

To examine how discussions evolve after an initial post, we compared the sentiment of each post with the sentiment of its associated comments using paired Wilcoxon signed-rank tests. This analysis was conducted at the aggregate regional level (United States and Europe) and within individual U.S. states and European countries. Only posts with at least one associated comment were included in the paired analysis.

At the regional level, both the United States and Europe exhibit statistically significant declines in sentiment from posts to comments (Table~\ref{tab:wilcoxon_region}). In the United States, the median post sentiment is 0.361, while the median comment sentiment decreases to 0.213, with the Wilcoxon signed-rank test indicating a highly significant difference ($W = 55{,}306{,}883$, $p < 10^{-83}$). In Europe, the decline is larger in magnitude, with the median post sentiment decreasing from 0.607 to 0.166 ($W = 18{,}216{,}016$, $p < 10^{-141}$).

State-level analyses within the United States further demonstrate that post--comment sentiment shifts vary across states (Table~\ref{tab:wilcoxon_states}). States such as Texas, Washington, Oregon, Colorado, and the District of Columbia show large and statistically significant declines in sentiment after false discovery rate correction. In contrast, several states with smaller sample sizes do not exhibit statistically significant differences after correction.

A similar pattern is observed across European countries (Table~\ref{tab:wilcoxon_countries}). Countries with large numbers of paired posts, including the United Kingdom, Germany, France, Denmark, and the Netherlands, exhibit highly significant post--comment sentiment declines. While the magnitude of the decline is directionally consistent across countries, statistical significance is reduced in countries with smaller sample sizes.

\begin{table}[!htbp]
\centering
\small
\caption{Paired comparison of post and comment sentiment at the regional level using the Wilcoxon signed-rank test.}
\label{tab:wilcoxon_region}
\setlength{\tabcolsep}{6pt}
\renewcommand{\arraystretch}{1.2}
\resizebox{\columnwidth}{!}{%
\begin{tabular}{lccccc}
\toprule
Region & $N$ (paired) & Median Post & Median Comment & $W$ & $p$-value \\
\midrule
United States & 16{,}731 & 0.361 & 0.213 & 55{,}306{,}883 & $1.72\times10^{-84}$ \\
Europe & 10{,}190 & 0.607 & 0.166 & 18{,}216{,}016 & $1.27\times10^{-142}$ \\
\bottomrule
\end{tabular}%
}
\end{table}

\begin{table}[!htbp]
\centering
\small
\caption{Paired Wilcoxon signed-rank test results comparing post and comment sentiment across U.S.\ states.}
\label{tab:wilcoxon_states}
\setlength{\tabcolsep}{4pt}
\renewcommand{\arraystretch}{1.15}
\resizebox{\columnwidth}{!}{%
\begin{tabular}{lcccccc}
\toprule
State & $N$ & Median Post & Median Comment & $W$ & $p$ & $p_{\text{FDR}}$ \\
\midrule
Texas & 3{,}096 & 0.348 & 0.235 & 1{,}707{,}476 & $6.28\times10^{-32}$ & $1.32\times10^{-30}$ \\
Washington & 1{,}376 & 0.439 & 0.205 & 316{,}706 & $5.49\times10^{-26}$ & $5.76\times10^{-25}$ \\
Oregon & 903 & 0.509 & 0.223 & 135{,}878 & $2.05\times10^{-16}$ & $1.44\times10^{-15}$ \\
Colorado & 890 & 0.440 & 0.278 & 145{,}870 & $1.28\times10^{-9}$ & $6.70\times10^{-9}$ \\
District of Columbia & 853 & 0.459 & 0.244 & 134{,}612 & $1.75\times10^{-9}$ & $7.34\times10^{-9}$ \\
\bottomrule
\end{tabular}%
}
\end{table}

\begin{table}[!htbp]
\centering
\small
\caption{Paired Wilcoxon signed-rank test results comparing post and comment sentiment across European countries.}
\label{tab:wilcoxon_countries}
\setlength{\tabcolsep}{4pt}
\renewcommand{\arraystretch}{1.15}
\resizebox{\columnwidth}{!}{%
\begin{tabular}{lcccccc}
\toprule
Country & $N$ & Median Post & Median Comment & $W$ & $p$ & $p_{\text{FDR}}$ \\
\midrule
United Kingdom & 4{,}201 & 0.511 & 0.145 & 3{,}503{,}967 & $6.22\times10^{-31}$ & $7.46\times10^{-30}$ \\
Germany & 1{,}699 & 0.637 & 0.162 & 492{,}729 & $1.06\times10^{-28}$ & $6.38\times10^{-28}$ \\
France & 480 & 0.847 & 0.313 & 24{,}088 & $2.30\times10^{-27}$ & $9.20\times10^{-27}$ \\
Denmark & 464 & 0.787 & 0.238 & 26{,}950 & $2.50\times10^{-20}$ & $7.50\times10^{-20}$ \\
Netherlands & 1{,}342 & 0.533 & 0.166 & 313{,}059 & $2.64\times10^{-19}$ & $6.34\times10^{-19}$ \\
\bottomrule
\end{tabular}%
}
\end{table}

\subsection{Geographic Variation in Sentiment Across U.S. States}

To assess whether cycling-related sentiment varies across U.S. states, we applied Kruskal--Wallis tests to both post-level sentiment scores and mean comment sentiment per post. This non-parametric approach was selected due to the non-normal distribution of sentiment scores and unequal sample sizes across states. Only states with at least 30 posts were included in the analysis.

Results of the Kruskal--Wallis tests are summarized in Table~\ref{tab:kw_us_states}. For post-level sentiment, the test indicates statistically significant differences across states ($H = 300.06$, $p < 10^{-49}$). However, the associated effect size is small ($\eta^2 = 0.013$), indicating that state-level differences explain a limited proportion of the overall variability in post sentiment.

A similar pattern is observed for mean comment sentiment per post. The Kruskal--Wallis test again reveals statistically significant differences across states ($H = 306.74$, $p < 10^{-52}$), with a slightly larger but still small effect size ($\eta^2 = 0.017$). In both cases, the number of states included in the analysis exceeds twenty, and total sample sizes exceed 16,000 observations, providing high statistical power to detect even small differences.

\begin{figure}[!htbp]
  \centering
  \includegraphics[
    width=0.85\columnwidth,
    height=0.35\textheight,
    keepaspectratio
  ]{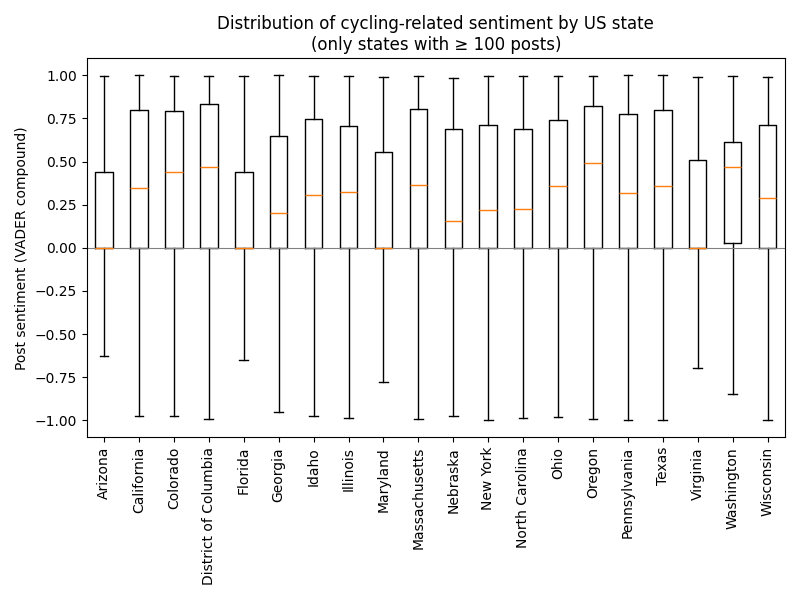}
  \caption{Distribution of post-level cycling sentiment across U.S. states with at least 30 posts.}
  \label{fig:us_state_boxplot}
\end{figure}

These results indicate that while cycling-related sentiment varies significantly across U.S. states, the magnitude of these differences is modest relative to the overall variability in sentiment scores. Figure~\ref{fig:us_state_boxplot} shows the distribution of post-level cycling sentiment across U.S. states included in the analysis.

\begin{table}[!htbp]
\centering
\small
\caption{Kruskal--Wallis tests for geographic variation in cycling-related sentiment across U.S.\ states.}
\label{tab:kw_us_states}
\setlength{\tabcolsep}{6pt}
\renewcommand{\arraystretch}{1.15}
\resizebox{\columnwidth}{!}{%
\begin{tabular}{lcccccc}
\toprule
Analysis & $H$ & $p$-value & $\eta^2$ & $k$ (states) & $N$ & Min.\ per state \\
\midrule
Post sentiment & 300.06 & $1.19\times10^{-50}$ & 0.013 & 23 & 21{,}107 & 30 \\
Mean comment sentiment & 306.74 & $3.40\times10^{-53}$ & 0.017 & 21 & 16{,}695 & 30 \\
\bottomrule
\end{tabular}%
}
\end{table}

\subsection{Geographic Variation in Sentiment Across European Countries}

We next examined whether cycling-related sentiment varies across European countries. Differences in post-level sentiment were assessed using the Kruskal--Wallis test. The analysis included 12 European countries and a total of 10{,}622 posts.

The Kruskal--Wallis test indicates statistically significant differences in post sentiment across countries ($H = 236.95$, $p = 1.50 \times 10^{-44}$; Table~\ref{tab:kw_europe}). The associated effect size is small ($\eta^2 = 0.021$), indicating that country-level differences account for a limited proportion of the overall variability in sentiment scores, despite the strong statistical significance.

\begin{table}[!htbp]
\centering
\small
\caption{Kruskal--Wallis tests for geographic variation in cycling-related sentiment across European countries.}
\label{tab:kw_europe}
\resizebox{\columnwidth}{!}{%
\begin{tabular}{lccccc}
\toprule
Analysis & $H$ & $p$-value & $\eta^2$ & $k$ (countries) & $N$ \\
\midrule
Post sentiment 
& 236.95 
& $1.50\times10^{-44}$ 
& 0.021 
& 12 
& 10{,}622 \\

Mean comment sentiment per post 
& 377.11 
& $4.39\times10^{-74}$ 
& 0.036 
& 12 
& 10{,}190 \\
\bottomrule
\end{tabular}%
}
\end{table}

\begin{figure*}[t!]
  \centering
  \includegraphics[width=0.9\textwidth]{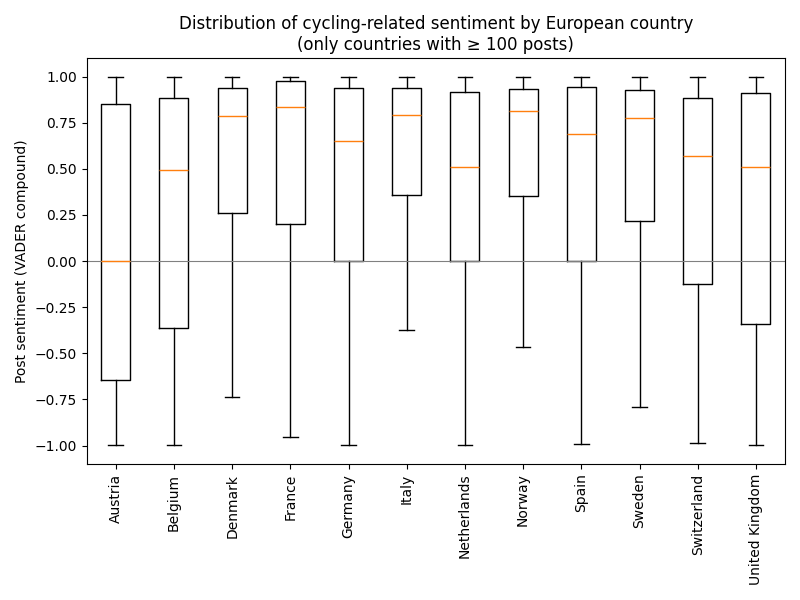}
  \caption{Distribution of post-level cycling sentiment across European countries with at least 100 posts.}
  \label{fig:eu_country_boxplot}
\end{figure*}

To identify which countries differ from one another, post-hoc pairwise comparisons were conducted using Dunn’s test with false discovery rate correction for multiple comparisons. The results of the Dunn post-hoc analysis are summarized in Table~\ref{tab:dunn_europe}. Significant differences are primarily observed between countries with higher median post sentiment, such as France, Denmark, Sweden, and Norway, and countries with lower median post sentiment, including Austria, Belgium, and the United Kingdom. In contrast, many pairwise comparisons among countries with similarly high sentiment levels are not statistically significant after correction, indicating clustering of countries with comparable sentiment distributions. Figure~\ref{fig:eu_country_boxplot} visualizes the distribution of post-level cycling sentiment across European countries included in the analysis.

\begin{table*}[!htbp]
\centering
\scriptsize
\caption{Dunn post-hoc pairwise comparisons for post-level sentiment across European countries (FDR-corrected $p$-values).}
\label{tab:dunn_europe}
\resizebox{\textwidth}{!}{%
\begin{tabular}{lcccccccccccc}
\toprule
 & AT & BE & DK & FR & DE & IT & NL & NO & ES & SE & CH & UK \\
\midrule
Austria (AT) & 1 & 0.360 & $1.7\times10^{-14}$ & $1.9\times10^{-26}$ & $1.5\times10^{-9}$ & $7.3\times10^{-8}$ & $2.2\times10^{-5}$ & $3.6\times10^{-11}$ & $1.0\times10^{-8}$ & $1.2\times10^{-10}$ & 0.137 & $6.7\times10^{-5}$ \\
Belgium (BE) & 0.360 & 1 & $3.4\times10^{-6}$ & $8.6\times10^{-15}$ & 0.069 & 0.0015 & 1 & $4.4\times10^{-5}$ & 0.0115 & $1.2\times10^{-4}$ & 1 & 1 \\
Denmark (DK) & $1.7\times10^{-14}$ & $3.4\times10^{-6}$ & 1 & 0.382 & 0.0219 & 1 & $4.1\times10^{-5}$ & 1 & 1 & 1 & 0.0178 & $4.3\times10^{-8}$ \\
France (FR) & $1.9\times10^{-26}$ & $8.6\times10^{-15}$ & 0.382 & 1 & $3.4\times10^{-11}$ & 1 & $7.6\times10^{-16}$ & 1 & 0.0026 & 1 & $2.5\times10^{-7}$ & $1.4\times10^{-22}$ \\
Germany (DE) & $1.5\times10^{-9}$ & 0.069 & 0.0219 & $3.4\times10^{-11}$ & 1 & 0.430 & 1 & 0.0666 & 1 & 0.161 & 1 & 0.0039 \\
Italy (IT) & $7.3\times10^{-8}$ & 0.0015 & 1 & 1 & 0.430 & 1 & 0.0241 & 1 & 1 & 1 & 0.0802 & 0.0033 \\
Netherlands (NL) & $2.2\times10^{-5}$ & 1 & $4.1\times10^{-5}$ & $7.6\times10^{-16}$ & 1 & 0.0241 & 1 & 0.00093 & 0.252 & 0.0026 & 1 & 1 \\
Norway (NO) & $3.6\times10^{-11}$ & $4.4\times10^{-5}$ & 1 & 1 & 0.0666 & 1 & 0.00093 & 1 & 1 & 1 & 0.0189 & $2.8\times10^{-5}$ \\
Spain (ES) & $1.0\times10^{-8}$ & 0.0115 & 1 & 0.0026 & 1 & 1 & 0.252 & 1 & 1 & 1 & 1 & 0.0103 \\
Sweden (SE) & $1.2\times10^{-10}$ & $1.2\times10^{-4}$ & 1 & 1 & 0.161 & 1 & 0.0026 & 1 & 1 & 1 & 0.0403 & $8.7\times10^{-5}$ \\
Switzerland (CH) & 0.137 & 1 & 0.0178 & $2.5\times10^{-7}$ & 1 & 0.0802 & 1 & 0.0189 & 1 & 0.0403 & 1 & 1 \\
United Kingdom (UK) & $6.7\times10^{-5}$ & 1 & $4.3\times10^{-8}$ & $1.4\times10^{-22}$ & 0.0039 & 0.0033 & 1 & $2.8\times10^{-5}$ & 0.0103 & $8.7\times10^{-5}$ & 1 & 1 \\
\bottomrule
\end{tabular}%
}
\end{table*}

\subsection{Mixed-Effects Modeling of City-Level Variation in the United States}

We estimated linear mixed-effects models for cycling-related sentiment in the United States. In these models, U.S.\ state was included as a fixed effect, while city was modeled as a random intercept to capture unobserved city-level heterogeneity. Separate models were estimated for post-level sentiment and for mean comment sentiment per post.

Table~\ref{tab:lmm_us} summarizes the variance components of the mixed-effects models. For post-level sentiment, the estimated city-level random intercept variance is 0.0027, compared to a residual variance of 0.2451. This corresponds to an intraclass correlation coefficient (ICC) of 0.011, indicating that approximately 1.1\% of the total variance in post sentiment is attributable to differences between cities.

For mean comment sentiment per post, the estimated city-level variance is slightly larger relative to the residual variance (0.0022 versus 0.0772), yielding an ICC of 0.028. This indicates that city-level differences account for approximately 2.8\% of the total variance in comment sentiment. In both models, the majority of sentiment variability is explained by within-city variation rather than between-city differences.

\begin{table}[!htbp]
\centering
\small
\caption{Variance components and intraclass correlation coefficients (ICC) from mixed-effects models of cycling-related sentiment in the United States. City is modeled as a random intercept, with state included as a fixed effect.}
\label{tab:lmm_us}
\setlength{\tabcolsep}{6pt}
\renewcommand{\arraystretch}{1.15}
\resizebox{\columnwidth}{!}{%
\begin{tabular}{lcccc}
\toprule
Outcome & City variance & Residual variance & ICC (city) & Optimizer \\
\midrule
Post sentiment & 0.0027 & 0.2451 & 0.0109 & Powell \\
Mean comment sentiment & 0.0022 & 0.0772 & 0.0275 & Powell \\
\bottomrule
\end{tabular}%
}
\end{table}

\subsection{Mixed-Effects Modeling of City-Level Variation in Europe}

Having identified differences in sentiment across regions, city-level variation in cycling-related sentiment was analyzed using linear mixed-effects models, with posts nested within cities and cities nested within countries. In these models, country was included as a fixed effect, while city was modeled as a random intercept to capture unobserved city-level heterogeneity. Separate models were estimated for post-level sentiment and for mean comment sentiment per post.

Table~\ref{tab:lmm_europe} reports the estimated variance components and intraclass correlation coefficients (ICC) for the European mixed-effects models. For post-level sentiment, the estimated city-level random intercept variance is 0.0068, compared to a residual variance of 0.4210. This corresponds to an ICC of 0.016, indicating that approximately 1.6\% of the total variance in post sentiment is attributable to differences between cities.

For mean comment sentiment per post, the estimated city-level variance is 0.0024, with a residual variance of 0.0632, yielding an ICC of 0.037. This indicates that approximately 3.7\% of the total variance in comment sentiment is explained by city-level differences. As in the United States, the majority of sentiment variability in Europe occurs within cities rather than between cities.

\begin{table}[!htbp]
\centering
\small
\caption{Variance components and intraclass correlation coefficients (ICC) from mixed-effects models of cycling-related sentiment in Europe. City is modeled as a random intercept, with country included as a fixed effect.}
\label{tab:lmm_europe}
\setlength{\tabcolsep}{6pt}
\renewcommand{\arraystretch}{1.15}
\resizebox{\columnwidth}{!}{%
\begin{tabular}{lcccc}
\toprule
Outcome & City variance & Residual variance & ICC (city) & Optimizer \\
\midrule
Post sentiment & 0.0068 & 0.4210 & 0.0158 & Powell \\
Mean comment sentiment & 0.0024 & 0.0632 & 0.0373 & Powell \\
\bottomrule
\end{tabular}%
}
\end{table}

\section{Discussion}
%paragraph zero: overall summary of what you did. 

This study introduces a framework for examining cycling perception through large-scale online discussions by combining sentiment analysis, topic modeling, aspect-based evaluation, and geographic comparison. The results show that sentiment toward cycling infrastructure is shaped more by discussion context and everyday experiential themes than by geographic differences alone. Across both regions, cycling discourse reflects largely shared experiences, with variation arising mainly from differences in emphasis rather than fundamentally different perspectives. Methodologically, the analysis demonstrates that distinguishing between posts and comments and linking sentiment to topics, aspects, and spatial scale provides a more nuanced understanding of how evaluation develops through online interaction beyond aggregate sentiment measures.

% Paragraph 1: Overall Patterns in Online Cycling Perception

Within the Reddit corpus analyzed in this study, sentiment toward cycling infrastructure varies more consistently with discussion content and structure than with geographic grouping. Across posts from the United States and the selected European countries, sentiment differs between original posts and their associated comment threads, with comments exhibiting a more critical tone. Although posts from the selected European countries exhibit slightly higher mean sentiment than those from the United States, the magnitude of this difference is small and the sentiment distributions overlap substantially. This pattern extends across geographic scales, where statistically significant differences emerge but contribute only modestly to overall variation. In the context of this dataset, these findings indicate that variation in online cycling sentiment is primarily associated with how specific cycling-related issues are discussed within conversations rather than with geographic location alone.

Our study's topic modeling shows that cycling discussions on Reddit are organized around a set of recurring, experience-based themes rather than general evaluations of cycling policy. Similar to the patterns observed in the present study, prior social media research on cycling infrastructure similarly finds that online discourse tends to center on concrete issues such as infrastructure design, safety-related events, construction impacts, and everyday use, indicating that discussion content is shaped primarily by lived experience rather than abstract attitudes \citep{ferster2021advocacy}. Studies using Twitter data further demonstrate that cycling-related conversations cluster around practical contexts, including daily routines, environmental conditions, and immediate challenges associated with riding, reflecting the event- and experience-oriented nature of online cycling communication \citep{das2019extracting,rahim2019public}.

While these findings align with earlier social media studies showing that cycling discussions focus on everyday experiences, the present analysis also highlights how these perceptions develop within discussion itself and vary across spatial contexts simultaneously. By examining posts and comment threads together and linking thematic patterns to geographic scale, the results move beyond identifying what people discuss toward understanding how evaluation emerges through interaction and local context. This perspective helps explain why broadly similar themes appear across regions even when overall sentiment differences remain small.

Within the present dataset, dominant topics across both the United States and Europe include bike theft and security, recreational and social cycling, route planning and commuting, and specific infrastructure elements such as bike lanes, paths, parking, and maintenance. While these themes appear in both regions, their relative prominence differs. Discussions in the United States more often emphasize commuting practices, group rides, route selection, and bike lane design, whereas European discussions place greater emphasis on place-based cycling contexts, maintenance and repair activities, parking and storage, and cycling-related travel experiences. These differences suggest that regional variation in discourse reflects shifts in emphasis within a shared experiential framework rather than fundamentally different ways of framing cycling. This pattern is consistent with broader mobility research showing that social media discourse tends to organize around practice-based thematic clusters rather than high-level sustainability narratives \citep{stiebe2024social,cebeci2023barriers}.

Sentiment varies systematically across the discussion topics identified through topic modeling. In both the United States and Europe, topics centered on recreational cycling, group rides and social cycling, route planning and commuting, and cycling-related travel experiences exhibit higher mean sentiment values, reflecting discussions framed around everyday use and mobility practices. Similar patterns are reported in prior social media research. \citep{anton2025trends} show that overall sentiment toward cycling on X platform (formerly Twitter) is predominantly positive, with positive perceptions linked to health, mobility, and recreational benefits. Using Twitter data from U.S. cities, \citep{das2019extracting} likewise report generally positive sentiment toward cycling that improves over time, particularly in discussions related to regular travel and commuting. In sustainable mobility contexts, \citep{torres2025using} find predominantly positive sentiment toward cycling as an active travel mode within university environments.

In contrast, topics related to bike theft and security are associated with the lowest mean sentiment values in both regions, indicating a predominantly negative emotional tone within discussions focused on safety and loss. Prior research similarly shows that concerns related to bicycle parking and theft can trigger strong negative emotional responses among cyclists, including anger, fear, and frustration \citep{meenar2019mapping}. Comparable patterns have been reported in prior social media studies. \citep{das2019extracting} associate negative sentiment in cycling discussions with crime, crashes, and other adverse conditions experienced by riders, highlighting how safety-related events contribute strongly to negative emotional expression online. Examining discussions of new cycling infrastructure,
\citep{lambert2026exploring} further show that sentiment surrounding cycle lanes and low-traffic neighborhoods is often polarized, with negative reactions linked to perceived safety concerns and implementation challenges.

Aspect-based sentiment patterns in Reddit discussions are consistent with prior social media studies showing generally positive attitudes toward cycling infrastructure that vary by infrastructure type rather than by location alone. In both the United States and Europe, higher sentiment is associated with paths and trails, transit integration, and general infrastructure, while lower sentiment concentrates around painted lanes, intersections and signals, and construction and roadworks. These aspect-level differences suggest that sentiment reflects how cyclists evaluate infrastructure through everyday riding experience and local expectations. Across both the United States and the selected European countries, protected lanes exhibit higher sentiment than painted lanes, an intuitive pattern given that physically separated infrastructure is generally associated with greater perceived safety and comfort for riders. In the United States, the more positive sentiment associated with protected lanes and general infrastructure likely reflects how physically separated facilities are discussed as noticeable improvements in safety and riding comfort. Such reactions may also be stronger in contexts where protected infrastructure is comparatively less common and therefore more salient in user experience. Discussions of transit integration exhibit positive sentiment in both regions, with the selected European countries showing higher sentiment, suggesting that transit integration is evaluated as part of routine mobility, with perceptions shaped primarily by usability and continuity. Discussions of construction and roadworks exhibit positive sentiment in both regions, with the selected European countries showing more positive evaluations, suggesting that construction activity is interpreted as an indicator of ongoing improvement to cycling infrastructure and associated more with development and progress than with disruption alone.Similar patterns have been observed in studies using X, where negative sentiment clusters around safety concerns, interactions with motor vehicles, and dissatisfaction with on-street or temporary facilities, even when overall support for cycling remains high \citep{das2019extracting,cebeci2023barriers,anton2025trends,lambert2026exploring}.

Cycling sentiment differs significantly between the United States and Europe, with European posts showing slightly higher average sentiment, although the magnitude of this difference is small and sentiment distributions overlap substantially. Similar patterns have been reported in transportation-related sentiment studies, where cross-national differences are statistically detectable, yet a large share of sentiment variation remains within regions and countries rather than between them \citep{zajacexploring,wang2024sentiments}. Sentiment also varies across U.S. states and European countries, but analyses at finer geographic scales show that much of this variation occurs within individual locations. City-level analyses further indicate substantial heterogeneity in sentiment within the same cities for both posts and comments.

Across both regions, sentiment in comment discussions is consistently less positive than in the original posts. This post--comment difference is statistically significant in both the United States and Europe and appears across states and countries. A similar pattern has been reported by \citep{ferster2020advocacy} in their analysis of protected bike lane discussions, where replies increasingly emphasized criticism even when initial posts were supportive. In transport-focused Twitter data, \citep{el2019linking} likewise show that sentiment becomes more negative as interactions progress, with later exchanges concentrating dissatisfaction. Evidence from metro and rail systems further suggests that replies and comment threads are dominated by complaints despite neutral or positive post-level sentiment \citep{agrawal2022implications,liu2025spatiotemporal}. The larger post--comment sentiment decline observed in Europe is therefore consistent with prior findings that discussion-level sentiment often diverges from initial framing, with comment threads expressing more critical evaluations than original posts \citep{ruan2023public}.

In practical terms, this study provides a structured approach that allows researchers and decision-makers to understand how the public perceives specific cycling-related topics and to identify where dissatisfaction is concentrated rather than relying solely on average sentiment measures. Topic- and sentiment-based analysis of online discussions can reveal recurring sources of criticism and help prioritize issues or locations for further investigation when negative sentiment clusters around particular themes or places. Linking sentiment to infrastructure-related aspects further enables identification of which elements of cycling infrastructure are most frequently criticized, providing clearer signals about where public concern is focused. The substantial variation observed within cities also allows results to guide more targeted qualitative investigation, rather than treating regional averages as fully representative. More broadly, the same analytical approach can be applied to other transport or infrastructure debates in which public perception develops through ongoing online discussion.

%In practical terms, this study provides a structured way for decision-makers to interpret large-scale online discussions about cycling infrastructure beyond simple measures of overall positivity or negativity. Rather than relying on average sentiment at the regional or national level, the approach makes it possible to identify which types of cycling experiences consistently generate support and which repeatedly attract criticism. This allows planners to see whether public concern centers on specific infrastructure elements, recurring safety issues, or everyday usability challenges, instead of assuming that sentiment reflects general attitudes toward cycling as a mode. The findings also suggest that regional averages can mask meaningful variation within cities. Even where overall sentiment appears similar across geographic units, dissatisfaction may cluster around particular discussion contexts or local experiences. Recognizing this helps avoid overgeneralizing from aggregate indicators and supports more targeted qualitative investigation, design review, or engagement efforts in areas where concerns are most concentrated. More broadly, the study demonstrates how ongoing digital discussion can be translated into interpretable signals about infrastructure perception, offering a scalable complement to traditional public input mechanisms.

\section{Limitations and Future Work}

Several limitations should be considered when interpreting the findings of this study. Because the analysis relies on social media data, the results should be interpreted with caution and viewed as complementary to traditional survey evidence. First, Reddit users are not representative of the general population, and the literature suggests that the platform tends to overrepresent younger, more technologically engaged individuals \citep{olteanu2019social,wang2019demographic}. As a result, the sentiment patterns observed here may not fully reflect broader public opinion toward cycling infrastructure.
Future research could address this limitation by integrating Reddit-based analyses with additional data sources, such as survey data, public meeting transcripts, or other social media platforms with different demographic compositions. Comparing sentiment patterns across multiple platforms would help assess the robustness of observed thematic and regional differences and clarify the extent to which online discourse aligns with broader population-level attitudes.

Second, the distribution of posts is highly uneven across cities, states, and countries. Regions with more active Reddit communities contribute disproportionately to the analysis, which may bias aggregate sentiment estimates despite the use of non-parametric tests and hierarchical models. While minimum sample thresholds were applied in geographic comparisons, residual imbalance remains a limitation. Future work could explore alternative weighting or normalization strategies to account for uneven posting activity across geographic units. For example, sentiment estimates could be adjusted based on population size, cycling mode share, or posting frequency to reduce the influence of highly active communities. In addition, expanding the temporal scope of data collection or incorporating additional geographic communities may help reduce sparsity in underrepresented locations and support more balanced cross-regional comparisons.

Third, sentiment analysis relies on automated language models that capture overall emotional tone but may miss nuanced context, particularly in technical discussions, sarcastic remarks, or policy-oriented debates. Although VADER is well suited for informal social media text, it cannot fully capture the complexity of human judgment or domain-specific language related to transportation planning. Future research could explore the use of domain-adapted or context-aware sentiment models trained on transportation-related corpora to better capture nuanced infrastructure discussions. Incorporating manual validation or hybrid human–machine coding strategies would also improve interpretability and allow for calibration of automated sentiment scores against expert judgment. Such extensions would strengthen confidence in sentiment estimates, particularly in discussions involving technical terminology or policy framing.

Fourth, this study focuses on perception data and does not directly link sentiment to objective measures of infrastructure quality, safety, or performance. As such, sentiment scores cannot be interpreted as direct indicators of infrastructure effectiveness or cycling outcomes, but rather as reflections of how infrastructure is discussed and perceived online. Future research could integrate sentiment patterns with objective infrastructure and safety indicators, such as bicycle crash statistics, facility type inventories, network density measures, or mode share data, to examine whether online discourse aligns with measurable conditions on the ground. Linking perception-based signals with performance metrics would help clarify where sentiment reflects documented infrastructure gaps and where it may instead reflect broader cultural or contextual factors.

Beyond addressing the limitations outlined above, future research can further extend the analytical scope of this study in several directions. First, city-level statistical comparisons within the United States and Europe could be expanded to identify local contexts where sentiment diverges most strongly, particularly when combined with spatial clustering or neighborhood-level analysis. Such extensions would allow more fine-grained examination of intra-urban variation in cycling discourse.

Second, incorporating geolocation references within posts could enable mapping sentiment to specific corridors, intersections, or infrastructure projects, strengthening alignment between online discussion and planning or design interventions. Moving from city-level aggregation to corridor-level mapping would improve the practical interpretability of discourse patterns.

Additional extensions include integrating multilingual sentiment models to better capture non-English discussions across European contexts and applying the framework across multiple time periods to examine how discourse evolves in response to infrastructure changes or policy shifts. Together, these extensions would deepen the spatial and temporal resolution of online perception analysis while strengthening its connection to real-world cycling conditions.

% Future work could address these limitations in several ways. First, city-level statistical comparisons within the United States and within Europe could be expanded to identify local contexts where sentiment diverges most strongly, particularly when combined with spatial clustering or neighborhood-level analysis. Second, incorporating geolocation references within posts could enable mapping sentiment to specific corridors, intersections, or projects, allowing closer alignment with planning and design decisions.

% Additional extensions include integrating multilingual sentiment models to better capture non-English discussions, combining Reddit data with other social media platforms to broaden demographic coverage, and validating sentiment patterns against external data sources such as cycling mode share, infrastructure density, or crash statistics. Together, these extensions would strengthen the link between online discourse and real-world cycling conditions.

\section{Conclusion}

This study investigated how cycling infrastructure is discussed and evaluated in Reddit communities across the United States and selected European countries using a multi-level analytical framework combining sentiment analysis, topic modeling, aspect-based classification, and hierarchical modeling. Overall sentiment toward cycling is generally positive in both regions; however, the results show that evaluations vary considerably depending on the themes and infrastructure aspects being discussed. Discussions related to recreational cycling, route planning, and protected or general infrastructure tend to be more positive, while conversations addressing theft, painted lanes, and construction are consistently more critical. Although differences across regions, states, and countries are statistically detectable, their magnitude remains modest, with most variation occurring within cities rather than between them. Taken together, these findings indicate that differences in sentiment are more closely related to discussion content and the structure of post and comment discussions than to geographic grouping alone. By connecting emotional tone with thematic and spatial context, this study highlights how large-scale online discussions can complement traditional transportation research by offering insight into how everyday mobility experiences are expressed and interpreted in practice.

\clearpage
% \bibliographystyle{elsarticle-harv}
% \bibliography{cas-refs}

\bibliographystyle{plainnat}
\bibliography{cas-refs}

@book{PucherBuehler2012,
  title={City Cycling},
  author={Pucher, J. and Buehler, R.},
  isbn={9780262304993},
  series={Urban and Industrial Environments},
  url={https://books.google.com/books?id=226mCyz9JaEC},
  year={2012},
  publisher={MIT Press}
}

@article{Handy2014,
  title={Promoting cycling for transport: research needs and challenges},
  author={Handy, Susan and Van Wee, Bert and Kroesen, Maarten},
  journal={Transport reviews},
  volume={34},
  number={1},
  pages={4--24},
  year={2014},
  publisher={Taylor \& Francis}
}

@book{Kitchin2014,
  title={The data revolution: Big data, open data, data infrastructures and their consequences},
  author={Kitchin, Rob},
  year={2014},
  publisher={Sage}
}

@article{Gil2012,
  title={Social media use for news and individuals' social capital, civic engagement and political participation},
  author={Gil de Z{\'u}{\~n}iga, Homero and Jung, Nakwon and Valenzuela, Sebasti{\'a}n},
  journal={Journal of computer-mediated communication},
  volume={17},
  number={3},
  pages={319--336},
  year={2012},
  publisher={Oxford University Press Oxford, UK}
}

@inproceedings{HuttoGilbert2014,
  title={Vader: A parsimonious rule-based model for sentiment analysis of social media text},
  author={Hutto, Clayton and Gilbert, Eric},
  booktitle={Proceedings of the international AAAI conference on web and social media},
  volume={8},
  number={1},
  pages={216--225},
  year={2014}
}

@article{ReimersGurevych2019,
  title={Sentence-bert: Sentence embeddings using siamese bert-networks},
  author={Reimers, Nils and Gurevych, Iryna},
  journal={arXiv preprint arXiv:1908.10084},
  year={2019}
}

@article{McInnes2017,
  title={hdbscan: Hierarchical density based clustering.},
  author={McInnes, Leland and Healy, John and Astels, Steve and others},
  journal={J. Open Source Softw.},
  volume={2},
  number={11},
  pages={205},
  year={2017}
}

@article{Grootendorst2022,
  title={BERTopic: Neural topic modeling with a class-based TF-IDF procedure},
  author={Grootendorst, Maarten},
  journal={arXiv preprint arXiv:2203.05794},
  year={2022}
}

@article{Aldred2018,
  title={Inequalities in self-report road injury risk in Britain: A new analysis of National Travel Survey data, focusing on pedestrian injuries},
  author={Aldred, Rachel},
  journal={Journal of Transport \& Health},
  volume={9},
  pages={96--104},
  year={2018},
  publisher={Elsevier}
}

@article{Heinen2010,
  title={Commuting by bicycle: an overview of the literature},
  author={Heinen, Eva and Van Wee, Bert and Maat, Kees},
  journal={Transport reviews},
  volume={30},
  number={1},
  pages={59--96},
  year={2010},
  publisher={Taylor \& Francis}
}

@article{Garrard2008,
  title={Promoting transportation cycling for women: the role of bicycle infrastructure},
  author={Garrard, Jan and Rose, Geoffrey and Lo, Sing Kai},
  journal={Preventive medicine},
  volume={46},
  number={1},
  pages={55--59},
  year={2008},
  publisher={Elsevier}
}

@article{Winters2011,
  title={Motivators and deterrents of bicycling: comparing influences on decisions to ride},
  author={Winters, Meghan and Davidson, Gavin and Kao, Diana and Teschke, Kay},
  journal={Transportation},
  volume={38},
  number={1},
  pages={153--168},
  year={2011},
  publisher={Springer}
}

@article{handy2014promoting,
  title={Promoting cycling for transport: research needs and challenges},
  author={Handy, Susan and Van Wee, Bert and Kroesen, Maarten},
  journal={Transport reviews},
  volume={34},
  number={1},
  pages={4--24},
  year={2014},
  publisher={Taylor \& Francis}
}

@misc{Fishman2016,
  title={Cycling as transport},
  author={Fishman, Elliot},
  journal={Transport Reviews},
  volume={36},
  number={1},
  pages={1--8},
  year={2016},
  publisher={Taylor \& Francis}
}

@article{piatkowski2019carrots,
  title={Carrots versus sticks: assessing intervention effectiveness and implementation challenges for active transport},
  author={Piatkowski, Daniel P and Marshall, Wesley E and Krizek, Kevin J},
  journal={Journal of Planning Education and Research},
  volume={39},
  number={1},
  pages={50--64},
  year={2019},
  publisher={SAGE Publications Sage CA: Los Angeles, CA}
}

@article{Willis2015,
  title={Cycling under influence: summarizing the influence of perceptions, attitudes, habits, and social environments on cycling for transportation},
  author={Willis, Devon Paige and Manaugh, Kevin and El-Geneidy, Ahmed},
  journal={International Journal of Sustainable Transportation},
  volume={9},
  number={8},
  pages={565--579},
  year={2015},
  publisher={Taylor \& Francis}
}

@article{Banerjee2022,
  title={Facilitating bicycle commuting beyond short distances: insights from existing literature},
  author={Banerjee, Apara and {\L}ukawska, Miros{\l}awa and Jensen, Anders Fjendbo and Haustein, Sonja},
  journal={Transport reviews},
  volume={42},
  number={4},
  pages={526--550},
  year={2022},
  publisher={Taylor \& Francis}
}

@article{StinsonBhat2003,
  title={Commuter bicyclist route choice: Analysis using a stated preference survey},
  author={Stinson, Monique A and Bhat, Chandra R},
  journal={Transportation research record},
  volume={1828},
  number={1},
  pages={107--115},
  year={2003},
  publisher={SAGE Publications Sage CA: Los Angeles, CA}
}

@article{Ferster2021,
  title={From advocacy to acceptance: Social media discussions of protected bike lane installations},
  author={Ferster, Colin and Laberee, Karen and Nelson, Trisalyn and Thigpen, Calvin and Simeone, Michael and Winters, Meghan},
  journal={Urban Studies},
  volume={58},
  number={5},
  pages={941--958},
  year={2021},
  publisher={SAGE Publications Sage UK: London, England}
}

@article{DuranRodas2020,
  title={Bike-sharing: the good, the bad, and the future: an analysis of the public discussion on Twitter},
  author={Duran-Rodas, David and Villeneuve, Dominic and Wulfhorst, Gebhard},
  journal={European journal of transport and infrastructure research},
  volume={20},
  number={4},
  pages={38--58},
  year={2020}
}

@article{Nelson2021,
  title={Crowdsourced data for bicycling research and practice},
  author={Nelson, Trisalyn and Ferster, Colin and Laberee, Karen and Fuller, Daniel and Winters, Meghan},
  journal={Transport reviews},
  volume={41},
  number={1},
  pages={97--114},
  year={2021},
  publisher={Taylor \& Francis}
}

@article{Blei2003,
  title={Latent dirichlet allocation},
  author={Blei, David M and Ng, Andrew Y and Jordan, Michael I},
  journal={Journal of machine Learning research},
  volume={3},
  number={Jan},
  pages={993--1022},
  year={2003}
}

@article{Proferes2021,
  title={Studying reddit: A systematic overview of disciplines, approaches, methods, and ethics},
  author={Proferes, Nicholas and Jones, Naiyan and Gilbert, Sarah and Fiesler, Casey and Zimmer, Michael},
  journal={Social Media+ Society},
  volume={7},
  number={2},
  pages={20563051211019004},
  year={2021},
  publisher={SAGE Publications Sage UK: London, England}
}

@article{Shirgaokar2021,
  title={Using twitter to investigate responses to street reallocation during COVID-19: Findings from the US and Canada},
  author={Shirgaokar, Manish and Reynard, Darcy and Collins, Damian},
  journal={Transportation Research Part A: Policy and Practice},
  volume={154},
  pages={300--312},
  year={2021},
  publisher={Elsevier}
}

@article{lock2020social,
  title={Social media as passive geo-participation in transportation planning--how effective are topic modeling \& sentiment analysis in comparison with citizen surveys?},
  author={Lock, Oliver and Pettit, Christopher},
  journal={Geo-Spatial Information Science},
  volume={23},
  number={4},
  pages={275--292},
  year={2020},
  publisher={Taylor \& Francis}
}

@article{Lock2020,
  title={Social media as passive geo-participation in transportation planning--how effective are topic modeling \& sentiment analysis in comparison with citizen surveys?},
  author={Lock, Oliver and Pettit, Christopher},
  journal={Geo-Spatial Information Science},
  volume={23},
  number={4},
  pages={275--292},
  year={2020},
  publisher={Taylor \& Francis}
}

@article{Chowdhury2023,
  title={Applications of text mining in the transportation infrastructure sector: a review},
  author={Chowdhury, Sudipta and Alzarrad, Ammar},
  journal={Information},
  volume={14},
  number={4},
  pages={201},
  year={2023},
  publisher={MDPI}
}

@article{Braun2025,
  title={Barriers to cycling, barriers to health equity: Disparities in perceived cycling environments in the US},
  author={Braun, Lindsay M},
  journal={Journal of Cycling and Micromobility Research},
  volume={4},
  pages={100066},
  year={2025},
  publisher={Elsevier}
}

@article{FrielWachholz2025,
  title={Cyclists’ perception of cycling infrastructure: The relationship between safety, comfort, and comprehensibility},
  author={Friel, David and Wachholz, Sina},
  journal={Transportation Research Part F: Traffic Psychology and Behaviour},
  volume={115},
  pages={103367},
  year={2025},
  publisher={Elsevier}
}

@article{DelclosAlio2024,
  title={Perceptions of potential cycling infrastructure in a low-cycling context: Evidence from a medium-sized urban area},
  author={Delcl{\`o}s-Ali{\'o}, Xavier and den Hoed, Wilbert},
  journal={International Journal of Sustainable Transportation},
  volume={18},
  number={12},
  pages={999--1011},
  year={2024},
  publisher={Taylor \& Francis}
}

@book{liu2012sentiment,
  title={Sentiment Analysis and Opinion Mining},
  author={Liu, B.},
  isbn={9781608458844},
  series={Synthesis digital library of engineering and computer science},
  url={https://books.google.com/books?id=Gt8g72e6MuEC},
  year={2012},
  publisher={Morgan \& Claypool}
}

@inproceedings{HuLiu2004,
  title={Mining and summarizing customer reviews},
  author={Hu, Minqing and Liu, Bing},
  booktitle={Proceedings of the tenth ACM SIGKDD international conference on Knowledge discovery and data mining},
  pages={168--177},
  year={2004}
}

@article{Cliff1993,
  title={Dominance statistics: Ordinal analyses to answer ordinal questions.},
  author={Cliff, Norman},
  journal={Psychological bulletin},
  volume={114},
  number={3},
  pages={494},
  year={1993},
  publisher={American Psychological Association}
}

@article{ferster2021advocacy,
  title={From advocacy to acceptance: Social media discussions of protected bike lane installations},
  author={Ferster, Colin and Laberee, Karen and Nelson, Trisalyn and Thigpen, Calvin and Simeone, Michael and Winters, Meghan},
  journal={Urban Studies},
  volume={58},
  number={5},
  pages={941--958},
  year={2021},
  publisher={SAGE Publications Sage UK: London, England}
}

@article{rahim2019public,
  title={Public opinion on dockless bike sharing: A machine learning approach},
  author={Rahim Taleqani, Ali and Hough, Jill and Nygard, Kendall E},
  journal={Transportation research record},
  volume={2673},
  number={4},
  pages={195--204},
  year={2019},
  publisher={SAGE Publications Sage CA: Los Angeles, CA}
}

@article{stiebe2024social,
  title={Social big data mining for the sustainable mobility and transport transition: findings from a large-scale cross-platform analysis},
  author={Stiebe, Michael},
  journal={European Transport Research Review},
  volume={16},
  number={1},
  pages={28},
  year={2024},
  publisher={Springer}
}

@article{cebeci2023barriers,
  title={Barriers and drivers for biking: What can policymakers learn from social media analytics?},
  author={Cebeci, Halil {\.I}brahim and G{\"u}ner, Samet and Arslan, Yusuf and Aydemir, Emrah},
  journal={Journal of Transport \& Health},
  volume={28},
  pages={101542},
  year={2023},
  publisher={Elsevier}
}

@article{anton2025trends,
  title={Trends in the discussion of cycling in urban environments: An X-based study},
  author={Ant{\'o}n-Gonz{\'a}lez, Laura and Pellicer-Chenoll, Maite and Villarrasa-Sapi{\~n}a, Israel and Toca-Herrera, Jos{\'e}-Luis and Gonz{\'a}lez, Luis-Mill{\'a}n and Dev{\'\i}s-Dev{\'\i}s, Jos{\'e}},
  journal={PLoS One},
  volume={20},
  number={11},
  pages={e0330616},
  year={2025},
  publisher={Public Library of Science San Francisco, CA USA}
}

@article{das2019extracting,
  title={Extracting patterns from Twitter to promote biking},
  author={Das, Subasish and Dutta, Anandi and Medina, Gabriella and Minjares-Kyle, Lisa and Elgart, Zachary},
  journal={IATSS research},
  volume={43},
  number={1},
  pages={51--59},
  year={2019},
  publisher={Elsevier}
}

@article{torres2025using,
  title={Using Sentiment Analysis to Study the Potential for Improving Sustainable Mobility in University Campuses},
  author={Torres, Ewerton Chaves Moreira and de Picado-Santos, Lu{\'\i}s Guilherme},
  journal={Sustainability},
  volume={17},
  number={14},
  pages={6645},
  year={2025},
  publisher={MDPI}
}

@article{lambert2026exploring,
  title={Exploring public discourse about new cycle lanes and low-traffic neighbourhoods using Twitter/X data},
  author={Lambert, Isabella Malet and Poortinga, Wouter and Potoglou, Dimitris and Xenias, Dimitrios},
  journal={Travel Behaviour and Society},
  volume={42},
  pages={101128},
  year={2026},
  publisher={Elsevier}
}

@article{ferster2020advocacy,
  title={From advocacy to acceptance: Social media discussions of protected bike lane installations},
  author={Ferster, Colin and Laberee, Karen and Nelson, Trisalyn and Thigpen, Calvin and Simeone, Michael and Winters, Meghan},
  journal={Urban Studies},
  volume={58},
  number={5},
  pages={941--958},
  year={2021},
  publisher={SAGE Publications Sage UK: London, England}
}

@article{zajacexploring,
  title={Exploring transport perceptions in London: A twitter-based analysis by country of residence},
  author={ZAJAC, Martin and HOR{\'A}K, Ji{\v{r}}{\'\i} and OSORIO-ARJONA, Joaqu{\'\i}n and KUKULIA{\v{C}}, Pavel}
}

@article{wang2024sentiments,
  title={Sentiments of rural US communities on electric vehicles and infrastructure: Insights from Twitter data},
  author={Wang, Ming and Zhao, Li and Cochran, Abigail L},
  journal={Sustainability},
  volume={16},
  number={11},
  pages={4871},
  year={2024},
  publisher={MDPI}
}

@article{el2019linking,
  title={Linking social, semantic and sentiment analyses to support modeling transit customers’ satisfaction: Towards formal study of opinion dynamics},
  author={El-Diraby, Tamer and Shalaby, Amer and Hosseini, Moein},
  journal={Sustainable Cities and Society},
  volume={49},
  pages={101578},
  year={2019},
  publisher={Elsevier}
}

@article{agrawal2022implications,
  title={Implications of a Twitter data-centred methodology for assessing commuters’ perceptions of the Delhi metro in India},
  author={Agrawal, Apoorv and Kuriakose, Paulose N},
  journal={Computational Urban Science},
  volume={2},
  number={1},
  pages={38},
  year={2022},
  publisher={Springer}
}

@article{liu2025spatiotemporal,
  title={A spatiotemporal analytical framework for dynamic public opinion and sentiment analysis on mega infrastructure project},
  author={Liu, Xishan and Shuang, Qing and Xu, XinXin and Zheng, Zhike},
  journal={Engineering, Construction and Architectural Management},
  pages={1--25},
  year={2025},
  publisher={Emerald Publishing Limited}
}

@article{ruan2023public,
  title={Public perception of electric vehicles on Reddit and Twitter: A cross-platform analysis},
  author={Ruan, Tao and Lv, Qin},
  journal={Transportation research interdisciplinary perspectives},
  volume={21},
  pages={100872},
  year={2023},
  publisher={Elsevier}
}

@article{blitz2021does,
  title={How does the individual perception of local conditions affect cycling? An analysis of the impact of built and non-built environment factors on cycling behaviour and attitudes in an urban setting},
  author={Blitz, Andreas},
  journal={Travel behaviour and society},
  volume={25},
  pages={27--40},
  year={2021},
  publisher={Elsevier}
}

@article{biassoni2023choosing,
  title={Choosing the bicycle as a mode of transportation, the influence of infrastructure perception, travel satisfaction and pro-environmental attitude, the case of milan},
  author={Biassoni, Federica and Lo Carmine, Chiara and Perego, Paolo and Gnerre, Martina},
  journal={Sustainability},
  volume={15},
  number={16},
  pages={12117},
  year={2023},
  publisher={MDPI}
}

@article{li2019mining,
  title={Mining public opinion on transportation systems based on social media data},
  author={Li, Dawei and Zhang, Yujia and Li, Cheng},
  journal={Sustainability},
  volume={11},
  number={15},
  pages={4016},
  year={2019},
  publisher={MDPI}
}

@mastersthesis{azimi2024investigating,
  title={Investigating the Effect of Road Characteristics on Pedestrian and Bike Crash Frequency},
  author={Azimi, Shiva},
  year={2024},
  school={University of Hawai'i at Manoa}
}

@article{chen2025influence,
  title={Influence of road environmental factors on traffic accidents involving vulnerable road users through negative binomial models},
  author={Chen, Ying and Tian, Yi and Ouyang, Zhaoheng and Zhu, Jiaxun},
  journal={PloS one},
  volume={20},
  number={2},
  pages={e0317601},
  year={2025},
  publisher={Public Library of Science San Francisco, CA USA}
}

@article{das2023leveraging,
  title={Leveraging Twitter data for sentiment analysis of transit user feedback: An NLP framework},
  author={Das, Adway and Prajapati, Abhishek Kumar and Zhang, Pengxiang and Srinath, Mukund and Ranjbari, Andisheh},
  journal={arXiv preprint arXiv:2310.07086},
  year={2023}
}

@article{alam2021identifying,
  title={Identifying public perceptions toward emerging transportation trends through social media-based interactions},
  author={Alam, Md Rakibul and Sadri, Arif Mohaimin and Jin, Xia},
  journal={Future Transportation},
  volume={1},
  number={3},
  pages={794--813},
  year={2021},
  publisher={MDPI}
}

@article{yang2024intangible,
  title={From intangible to tangible: The role of big data and machine learning in walkability studies},
  author={Yang, Jun and Fricker, Pia and Jung, Alexander},
  journal={Computers, Environment and Urban Systems},
  volume={109},
  pages={102087},
  year={2024},
  publisher={Elsevier}
}

@article{olteanu2019social,
  title={Social data: Biases, methodological pitfalls, and ethical boundaries},
  author={Olteanu, Alexandra and Castillo, Carlos and Diaz, Fernando and K{\i}c{\i}man, Emre},
  journal={Frontiers in big data},
  volume={2},
  pages={13},
  year={2019},
  publisher={Frontiers Media SA}
}

@inproceedings{wang2019demographic,
  title={Demographic inference and representative population estimates from multilingual social media data},
  author={Wang, Zijian and Hale, Scott and Adelani, David Ifeoluwa and Grabowicz, Przemyslaw and Hartman, Timo and Fl{\"o}ck, Fabian and Jurgens, David},
  booktitle={The world wide web conference},
  pages={2056--2067},
  year={2019}
}

@article{meenar2019mapping,
  title={Mapping the emotional experience of travel to understand cycle-transit user behavior},
  author={Meenar, Mahbubur and Flamm, Bradley and Keenan, Kevin},
  journal={Sustainability},
  volume={11},
  number={17},
  pages={4743},
  year={2019},
  publisher={MDPI}
}

\clearpage
\appendix
\section{Subreddits Included in the Analysis}
\label{app:subreddits}

\subsection{United States Subreddits}
\begingroup
\small
\setlength{\LTleft}{\fill}
\setlength{\LTright}{\fill}
\setlength{\tabcolsep}{4pt} % tighter columns (optional)
\renewcommand{\arraystretch}{1.05} % a bit more row spacing (optional)

\setlength{\tabcolsep}{3pt} % default is 6pt; smaller = fits in margins
\small                      % optional but helps
\begin{longtable}{@{}p{3.2cm} p{3.0cm} p{5.5cm} r@{}}
\caption{Cycling-related subreddits from the United States included in the dataset. Subscriber counts reflect approximate values at the time of data collection.}
\label{tab:us_subreddits} \\

\toprule
\textbf{Subreddit} & \textbf{Title} & \textbf{Description} & \textbf{Subscribers} \\
\midrule
\endfirsthead

\multicolumn{4}{l}{\textit{Table~\ref{tab:us_subreddits} continued.}} \\
\toprule
\textbf{Subreddit} & \textbf{Title} & \textbf{Description} & \textbf{Subscribers} \\
\midrule
\endhead

\bottomrule
\endfoot

AtlantaCycling & Cycling in Atlanta, GA & Dedicated to discussion and information about riding a bicycle in the metro Atlanta area. & 73 \\
bicycling412 & Pittsburgh Bicycling & Subreddit for bicyclists in and around Pittsburgh, Pennsylvania. & 5{,}799 \\
BikeATL & BikeATL & Your Reddit resource for all things cycling in the Atlanta Metro area. & 1{,}275 \\
bikebmore & Bicycles in Baltimore & People riding bicycles in Baltimore. & 1{,}341 \\
BikeChicago & Biking in Chicago & A place for cycling events and information in Chicago, including photos, routes, and infrastructure discussions. & 415 \\
BikeCLE & BikeCLE & The cycling community of the Greater Cleveland Area. & 1{,}155 \\
bikedc & Cycling in the DC/MD/VA Area & Information and discussion for bicycling in the greater DC, Maryland, and Virginia metro area. & 13{,}154 \\
BikeGrandRapids & Cycling in Grand Rapids, MI & Community for cyclists in the Grand Rapids area. & 936 \\
bikehouston & Cycling in Houston, TX & A place for cyclists of all kinds in the Houston area. & 4{,}593 \\
BikeJax & Cycling in Jacksonville, FL & Local hub for bicycling events, routes, and advocacy in Jacksonville. & 148 \\
BikeLA & BikeLA & Community for cyclists in Los Angeles. & 15{,}113 \\
BikeOklahoma & Oklahoma Bicycling & Community for bicycling in Oklahoma. & 71 \\
BikePortland & Portland Cycling & Portland Oregon related biking discussions. & 291 \\
BikeSanDiego & Bike San Diego & Intersection of r/SanDiego and r/Bicycling discussions. & 362 \\
bikesatx & BikesATX & Biking in San Antonio, Texas. & 1{,}782 \\
BikeTulsa & Bike Tulsa & Cycling enthusiasts in Tulsa, Oklahoma. & 2 \\
BikingATX & BikingATX & Subreddit for cyclists in Austin, Texas. & 18{,}974 \\
BikingLasVegas & Las Vegas Cycling & Cycling discussions in and around the Las Vegas Valley. & 138 \\
BikingMad & Madison Area Biking & Cycling community for Madison, Wisconsin. & 1{,}632 \\
boisebike & Boise Bike & Community for bicycle enthusiasts in Boise. & 1{,}095 \\
BuffaloBike & Buffalo and WNY Bicycling & Cycling news and events in Buffalo and Western New York. & 887 \\
CarFreeChicago & CarFreeChicago & Advocacy for car-free and multimodal infrastructure in Chicago. & 7{,}127 \\
CarFreePhoenix & CarFreePhoenix & Community advocating reduced car use in Phoenix. & 491 \\
CarIndependentANC & CarIndependentANC & Car-independent and bike-friendly advocacy in Anchorage. & 72 \\
CarIndependentPGH & CarIndependentPGH & Advocacy-focused community for a more bike- and pedestrian-friendly Pittsburgh. & 744 \\
charlotte\_cyclists & Charlotte Cyclists & Cycling community in Charlotte, North Carolina. & 41 \\
chattanoogagravel & Chattanooga Gravel Cycling & Gravel and adventure cycling in the Chattanooga area. & 39 \\
chibike & Chicago Bicycling & Large cycling community in Chicago. & 20{,}742 \\
ChiGroupRides & Group Bicycle Rides in Chicago! & We're a few guys who ride every Tuesday night, but want others to come with, or to share when your group rides happen! Tentatively, we're riding the LFT every Tuesday night, starting at 6pm. Meetup location TBA. Bring your road bike, your hybrid, your fixie... We're a no-drop ride, open to riders of all skill level! & 144 \\

CincinnatiRiders & Cincinnati Motorcycle Riders & Cincinnati metropolitan area motorcycle riders. Organizing group rides, meetups and activities for all types of riders. & 387 \\

cincybiking & Cincinnati Bicycling & A bicycling subreddit for the Cincinnati area. & 910 \\

CLT\_Cyclists & Charlotte Cyclists & A subreddit dedicated to the cycling scene in Charlotte, North Carolina, the Queen City itself! This is a subreddit where we discuss the varying cycling communities, secret cycling spots, daily grievances of cycling, and so on, all within the city of Charlotte. Neighboring communities are also welcome! We are also the co-hosts of Charlotte's Critical Mass Ride! & 1{,}224 \\

COBike & Bicycling with Altitude & Bicycling in Colorado. Group rides, bike shop recommendations, news, and events! & 6{,}880 \\

ColumbusCycling & ColumbusCycling & Events and discussion regarding all things cycling related in Columbus and central Ohio. Please visit the Wiki for links and info. & 83 \\

CycleDenver & CycleDenver: Bicycling in Denver, Colorado & All things bicycling related in Denver, Colorado. & 82 \\

CycleDetroit & CycleDetroit & This is a place to discuss cycling in Detroit. This subreddit was created to replace r/BikeDetroit, which was locked in 2022. & 79 \\

CyclePDX & CyclePDX: Bicycling in Portland, Oregon & For all things related to bicycling in Portland Oregon. & 6{,}364 \\

CycleSeattle & CycleSeattle: Bicycling in Seattle, Washington & For all things related to bicycling in Seattle, Washington. & 223 \\

dfwbike & DFWbike - Cycling in Dallas/Ft. Worth & Dallas/Fort Worth has a broad and diverse bicycling culture. Discuss bicycling news, stories, and events in North Texas. & 3{,}680 \\

erideslascruces & erideslascruces & A place for personal electric vehicle riders in Las Cruces, NM, USA. Share pictures, meet up with people, discuss maintenance issues, have fun in the sun! & 48 \\

htowncyclistsgonewild & htowncyclistsgonewild & What did you expect? Post your bike, cool/odd things you see on your ride, or little known trails in the Houston area. & 55 \\

KnoxvilleBikes & Knoxville Bikes! & Doesn't matter whether you like road races, mountain biking, cyclocross, fixies, BMX, bike commuting, or anything in between. If you live/work/play in Knoxville or East Tennessee and you like to ride bikes, c'mon in and speak your mind. & 79 \\

miamibiking & Miami Biking & A friendly subreddit for all species of bicycle riders in Miami, Florida. We also play tag! & 766 \\

Micromobility\_ATL & Micromobility\_ATL & For all riders of bikes, scooters, personal electric vehicles, and anything else you can ride in the bike lanes, on the BeltLine, or on the streets around Atlanta. Check the Sidebar for more info. & 620 \\

MiltownBiking & Milwaukee's Bicycle Community & Welcome to Milwaukee's bike subreddit! From the urban commuters to the beach cruisers, everyone and their bike is welcome here for newbie advice, pro events, and everything in between! Bike maps and bike shops are listed in the wiki. & 1{,}857 \\

NYCbike & NYCbike: Cycling news and info for NYC & A resource for NYC-specific cycling events and information. Bike news that is not relevant to the New York area should be posted to /r/bicycling or /r/cycling instead. This is a great place to post and find group rides, questions about NYC cycling and bike shops, infrastructure changes, and cycling-related news. New to riding in the city? We'd love to help you get started! & 53{,}432 \\

OmahaMetroCycling & Omaha Metro Cycling & A place for Omaha/Council Bluffs area cyclists to share pictures, routes, group rides and anything else bicycle related. & 505 \\

phillycycling & Philly Cycling: Rides / Events / Shops / Chat & Philly Cycling: A Reddit Community dedicated to biking in Philly and the Greater Philadelphia Area. Feel free to post public bike rides, events, shops or just chat. & 9{,}514 \\

PhillyWandrers & Philly Wandrers & This community is for people who want to explore and experience Philly by bike. Rides are organized and led by the Bicycle Club of Philadelphia. & 47 \\

riddeit & Reddit bicycle rides around the Columbus area & Bicycle riding for Redditors in the Columbus, Ohio area. & 1{,}294 \\

rocbike & Bicycling Culture in Rochester, NY, US & A resource for Rochester, NY bicycle specific events/information. This is a good place to post about group rides, questions about Rochester bike shops, specific Rochester biking news (like a bike lane closure or opening), bicycling related political events/news/meetings, etc. & 340 \\

RVAbikes & Richmond Bicycle Forum & Online community for RVA cyclists. & 532 \\

seattlebike & bikes for seattleites & A resource for Seattle bicycle specific events and information, and a place for the Seattle bicycling community to gather. This is a good place to post about group rides, questions about Seattle bike shops, Seattle biking news (like a bike lane closure or opening), bicycling related political events/news/meetings, etc. & 11{,}289 \\

socialcyclingaustin & socialcyclingaustin & Established in 2009. Subscribe to our pages for more information on rides and events: www.facebook.com/socialcyclingaustin ; www.instagram.com/socialcyclingaustin. Social Cycling Austin brings Austin cyclists together to share our love of riding bikes. Our unique events promote socializing between all types of cyclists no matter what your skill level, what kind of bike you ride, or what terrain you like to tear up. Join us! & 706 \\

StPeteCycling & StPeteCycling & A subreddit for cycling in St Petersburg Florida. & 190 \\

TacomaCyclists & TacomaCyclists & A place for anyone riding a bicycle in Tacoma. & 138 \\

tampabikes & Tampa Bikes! & Tampa Bikes! & 128 \\
\end{longtable}

\endgroup

\subsection{European City Subreddits}
\begingroup
\small
\setlength{\LTleft}{\fill}
\setlength{\LTright}{\fill}
\setlength{\tabcolsep}{4pt}
\renewcommand{\arraystretch}{1.05}

\setlength{\tabcolsep}{3pt} % reduce column padding (default is 6pt)
\small                      % slightly smaller font
\begin{longtable}{@{}p{3.0cm} p{3.4cm} p{5.8cm} r@{}}
\caption{City-based European subreddits included in the dataset. Subscriber counts reflect approximate values at the time of data collection.}
\label{tab:eu_subreddits} \\

\toprule
\textbf{Subreddit} & \textbf{Title} & \textbf{Description} & \textbf{Subscribers} \\
\midrule
\endfirsthead

\multicolumn{4}{l}{\textit{Table~\ref{tab:eu_subreddits} continued.}} \\
\toprule
\textbf{Subreddit} & \textbf{Title} & \textbf{Description} & \textbf{Subscribers} \\
\midrule
\endhead

\bottomrule
\endfoot

Aalborg & Aalborg -- Dobbelt A & En subreddit for Aalborg. & 23{,}366 \\
Aarhus & Aarhus & Subreddit for those living in or near Aarhus, sharing local news, discussions, and meetups. & 38{,}532 \\
Amsterdam & Amsterdam & Subreddit for everything related to Amsterdam, in Dutch or English. & 348{,}091 \\
Antwerpen & 't Stad & Subreddit about the Belgian city of Antwerp. & 26{,}638 \\
Barcelona & Barcelona & Community for people living in Barcelona. & 161{,}321 \\
Bergen & Bergen, a city in Norway & All things related to Bergen, a city in Norway. & 29{,}091 \\
berlin & Neues aus Berlin & Bilingual subreddit for everything relating to Berlin, Germany. & 587{,}764 \\
berlinsocialclub & Berlin Social Club & Community for arranging meetups and social events in Berlin. & 100{,}532 \\
brum & Birmingham UK & Subreddit for people in Birmingham and the wider West Midlands. & 150{,}673 \\
brussels & Brussels & Subreddit for people living in or interested in Brussels. & 266{,}179 \\
cologne & Cologne -- K\"oln & All about the city of Cologne in Germany. & 119{,}695 \\
copenhagen & Copenhagen & Subreddit for all things Copenhagen. & 169{,}582 \\
Edinburgh & Edinburgh: News, Events and Discussion & Community for discussion related to Edinburgh. & 292{,}443 \\
eindhoven & Eindhoven & Community hub for questions and discussions about Eindhoven. & 51{,}509 \\
ExpatsTheHague & Expats in The Hague & News, chat, and advice for expats living in The Hague. & 716 \\
frankfurt & Frankfurt am Main & News and discussions for Frankfurt and the Rhein--Main area. & 420{,}088 \\
glasgow & Let Glasgow Flourish & Subreddit for everything Glasgow and the West. & 262{,}505 \\
graz & Graz -- Stadt der Ger\"ausche & Discussions related to the city of Graz, Austria. & 96{,}348 \\
hamburg & Hansestadt Hamburg & Subreddit for everything related to Hamburg, Germany. & 114{,}162 \\
london & London, UK & Subreddit for everyday London life and visitors. & 1{,}465{,}417 \\
londoncycling & London Cycling & Community for London-based cyclists sharing routes and local information. & 73{,}256 \\
LondonSocialClub & London Social Club & Social meetups and events in London. & 219{,}270 \\
Madrid & Madrid & Subreddit for Madrid and its surrounding region. & 177{,}221 \\
MadridsRedditBikeClub & Madrid Reddit Bike Club & Bike club for residents and visitors in Madrid. & 51 \\
Malmoe & Malmö & Subreddit for everything related to Malmö, Sweden. & 47{,}097 \\
manchester & Manchester & Community for Greater Manchester, UK. & 418{,}356 \\
Munich & Munich & Subreddit for residents and visitors of Munich. & 369{,}794 \\
oslo & Oslo & Subreddit for people living in or visiting Oslo. & 88{,}034 \\
paris & Paris et sa banlieue & Everything about Paris and Greater Paris. & 403{,}968 \\
ParisTravelGuide & Paris Travel Guide & Travel-focused subreddit for visitors to Paris. & 105{,}970 \\
rome & Rome, Italy & Subreddit for the city of Rome, ancient and modern. & 145{,}979 \\
Rotterdam & Rotterdam & News and information about Rotterdam, the Netherlands. & 154{,}280 \\
Salzburg & Salzburg & Subreddit for the city of Salzburg. & 30{,}674 \\
SocialParis & Paris Social Club & Meetups and social events in Paris. & 102{,}819 \\
stockholm & Stockholm & Subreddit dedicated to Stockholm and its surroundings. & 152{,}859 \\
TheHague & The Hague & News and information about The Hague. & 86{,}783 \\
TowerHamlets & Tower Hamlets & News and discussions related to Tower Hamlets, London. & 939 \\
trondheim & Trondheim & Questions and discussions about Trondheim. & 23{,}414 \\
Utrecht & Utrecht & Subreddit for the city of Utrecht. & 133{,}264 \\
valencia & Valencia & Subreddit for questions, meetups, and news about Valencia. & 94{,}730 \\
wien & Wien ist anders & Subreddit for residents and visitors of Vienna. & 209{,}432 \\
zurich & Zürich & Subreddit dedicated to Zürich and its greater area. & 123{,}081 \\

\end{longtable}
\endgroup

\section{Predefined Geographic Reference Cities}
\label{app:cities}

\subsection{United States Reference Cities Used for Data Collection}

Table~\ref{tab:us_cities} lists the predefined U.S.\ cities used as geographic reference anchors during data collection. Not all reference cities necessarily yielded posts in the final dataset.

\begin{longtable}{p{4cm} p{9cm}}
\caption{U.S.\ cities used as geographic references during data collection, grouped by state.}
\label{tab:us_cities}\\

\toprule
\textbf{State} & \textbf{Cities} \\
\midrule
\endfirsthead

\multicolumn{2}{l}{\textit{Table~\ref{tab:us_cities} continued.}} \\
\toprule
\textbf{State} & \textbf{Cities} \\
\midrule
\endhead

\bottomrule
\endfoot

Alabama & Huntsville; Birmingham; Montgomery; Mobile; Tuscaloosa \\
Alaska & Anchorage municipality; Fairbanks; Juneau city and borough; Wasilla; Sitka city and borough \\
Arizona & Phoenix; Tucson; Mesa; Chandler; Gilbert town \\
Arkansas & Little Rock; Fayetteville; Fort Smith; Springdale; Jonesboro \\
California & Los Angeles; San Diego; San Jose; San Francisco; Fresno \\
Colorado & Denver; Colorado Springs; Aurora; Fort Collins; Lakewood \\
Connecticut & Bridgeport; New Haven; Stamford; Hartford; Waterbury \\
Delaware & Wilmington; Dover; Newark; Middletown; Smyrna town \\
District of Columbia & Washington \\
Florida & Jacksonville; Miami; Tampa; Orlando; St.\ Petersburg \\
Georgia & Atlanta; Augusta-Richmond County consolidated government (balance); Columbus; Macon-Bibb County; Savannah \\
Hawaii & Urban Honolulu \\
Idaho & Boise City; Meridian; Nampa; Idaho Falls; Pocatello \\
Illinois & Chicago; Aurora; Joliet; Naperville; Rockford \\
Indiana & Indianapolis city (balance); Fort Wayne; Evansville; South Bend; Carmel \\
Iowa & Des Moines; Cedar Rapids; Davenport; Sioux City; Iowa City \\
Kansas & Wichita; Overland Park; Kansas City; Olathe; Topeka \\
Kentucky & Louisville/Jefferson County metro government (balance); Lexington-Fayette urban county; Bowling Green; Owensboro; Covington \\
Louisiana & New Orleans; Baton Rouge; Shreveport; Lafayette; Lake Charles \\
Maine & Portland; Lewiston; Bangor; South Portland; Auburn \\
Maryland & Baltimore; Frederick; Gaithersburg; Rockville; Bowie \\
Massachusetts & Boston; Worcester; Springfield; Cambridge; Lowell \\
Michigan & Detroit; Grand Rapids; Warren; Sterling Heights; Ann Arbor \\
Minnesota & Minneapolis; St.\ Paul; Rochester; Duluth; Bloomington \\
Mississippi & Jackson; Gulfport; Southaven; Hattiesburg; Biloxi \\
Missouri & Kansas City; St.\ Louis; Springfield; Columbia; Independence \\
Montana & Billings; Missoula; Great Falls; Bozeman; Butte-Silver Bow \\
Nebraska & Omaha; Lincoln; Bellevue; Grand Island; Kearney \\
Nevada & Las Vegas; Henderson; Reno; North Las Vegas; Sparks \\
New Hampshire & Manchester; Nashua; Concord; Derry; Dover \\
New Jersey & Newark; Jersey City; Paterson; Elizabeth; Lakewood township \\
New Mexico & Albuquerque; Las Cruces; Rio Rancho; Santa Fe; Roswell \\
New York & New York; Buffalo; Rochester; Yonkers; Syracuse \\
North Carolina & Charlotte; Raleigh; Greensboro; Durham; Winston-Salem \\
North Dakota & Fargo; Bismarck; Grand Forks; Minot; West Fargo \\
Ohio & Columbus; Cleveland; Cincinnati; Toledo; Akron \\
Oklahoma & Oklahoma City; Tulsa; Norman; Broken Arrow; Edmond \\
Oregon & Portland; Salem; Eugene; Gresham; Hillsboro \\
Pennsylvania & Philadelphia; Pittsburgh; Allentown; Reading; Erie \\
Rhode Island & Providence; Warwick; Cranston; Pawtucket; East Providence \\
South Carolina & Charleston; Columbia; North Charleston; Mount Pleasant town; Rock Hill \\
South Dakota & Sioux Falls; Rapid City; Aberdeen; Brookings; Watertown \\
Tennessee & Memphis; Nashville-Davidson metropolitan government (balance); Knoxville; Chattanooga; Clarksville \\
Texas & Houston; San Antonio; Dallas; Austin; Fort Worth \\
Utah & Salt Lake City; West Valley City; Provo; West Jordan; Orem \\
Vermont & Burlington; South Burlington; Rutland; Barre city; Montpelier \\
Virginia & Virginia Beach; Chesapeake; Norfolk; Richmond; Newport News \\
Washington & Seattle; Spokane; Tacoma; Vancouver; Bellevue \\
West Virginia & Charleston; Huntington; Morgantown; Parkersburg; Wheeling \\
Wisconsin & Milwaukee; Madison; Green Bay; Kenosha; Racine \\
Wyoming & Cheyenne; Casper; Gillette; Laramie; Rock Springs \\

\end{longtable}

\subsection{European Reference Cities Used for Data Collection}

Table~\ref{tab:european_cities} lists the predefined European cities used as geographic reference anchors during data collection. Not all reference cities necessarily yielded posts in the final dataset.

\begin{table}[!htbp]
\centering
\caption{European cities used as geographic references during data collection, grouped by country.}
\label{tab:european_cities}
\begin{tabular}{p{4cm} p{9cm}}
\toprule
\textbf{Country} & \textbf{Cities} \\
\midrule
United Kingdom & London; Birmingham; Manchester; Glasgow; Edinburgh \\
Germany & Berlin; Hamburg; Munich; Cologne; Frankfurt \\
France & Paris; Marseille; Lyon; Nice \\
Netherlands & Amsterdam; Rotterdam; The Hague; Utrecht; Eindhoven \\
Denmark & Copenhagen; Aarhus; Aalborg \\
Spain & Madrid; Barcelona; Valencia; Seville; Zaragoza \\
Italy & Rome; Milan; Naples; Turin \\
Sweden & Stockholm; Malmö \\
Norway & Oslo; Bergen; Trondheim \\
Austria & Vienna; Graz; Salzburg \\
Belgium & Brussels; Antwerp; Ghent \\
Switzerland & Zurich; Geneva; Basel \\
\bottomrule
\end{tabular}
\end{table}

\end{document}